\documentclass[acmsmall,screen,authorversion]{acmart}

\AtBeginDocument{%
  }

\setcopyright{none}
\usepackage{graphicx}
\usepackage{multirow}
\usepackage{makecell}
\usepackage{tcolorbox}
\usepackage{enumerate}
\usepackage[shortlabels]{enumitem}

\begin{document}

\title{Newer and Bigger, but Safer? A Longitudinal Study of the Functionality-Security Gap in LLM-Generated Code}

\author{Thiago Santos de Moura}
\affiliation{%
  \institution{Ruhr-Universität Bochum}
  \city{Bochum}
  \country{Germany}}
\email{thiago.santosdemoura@acm.org}
\orcid{0009-0007-1927-8378}

\author{Fynn Matuschek}
\affiliation{%
  \institution{Ruhr-Universität Bochum}
  \city{Bochum}
  \country{Germany}}
\email{fynn.matuschek@rub.de}
\orcid{0009-0003-1330-0264}

\author{Flavio Toffalini}
\affiliation{%
  \institution{Ruhr-Universität Bochum}
  \city{Bochum}
  \country{Germany}}
\email{flavio.toffalini@rub.de}
\orcid{0000-0002-7114-5640}

\author{Yannic Noller}
\authornote{Corresponding author: yannic.noller@acm.org}
\affiliation{%
  \institution{Ruhr-Universität Bochum}
  \city{Bochum}
  \country{Germany}}
\email{yannic.noller@acm.org}
\orcid{0000-0002-9318-8027}

\begin{abstract}
%%
% Problem
Large Language Models (LLMs) are widely used to generate code. Although their functional plausibility keeps improving, the generated code often contains security vulnerabilities. The functionality-security gap captures code that passes functional tests but fails security tests. 
%%
% Related Work Limitations
A recent longitudinal study of three model families concluded that LLMs become smarter but not safer, with the only considered open-weight family stagnating. Whether this holds for other (open-weight) families and particularly for compact models remains open.
%%
% Approach/Methodology
We present a longitudinal study of the gap across 32 LLMs from seven model families (five open-weight), covering three successive releases per family in flagship and compact variants. Using CWEval with 119 tasks in five programming languages and 31 CWEs, we compare trajectories across families, model sizes, and languages.
%%
% Results
Newer models do become safer in absolute terms, although no family closes the gap. Unlike prior work, we find that openness does not separate the families: every considered open-weight family narrows the gap significantly, while Gemini 3.1 Pro keeps a gap as wide as the one reported for Llama. Compact models usually produce less secure code than their flagship counterparts, with notable exceptions (e.g., Gemini 3.7 Flash). At the CWE level, we confirm persistent weaknesses such as log injection (CWE-117) and HTTP response splitting (CWE-113) and regressions in the newest proprietary models on memory and integer weaknesses, and show that the same CWE carries very different risk across languages.
%%
% Impact
From these results, we derive implications for LLM vendors, researchers, and developers. In particular, developers should assume neither that upgrades improve security nor that proprietary models are more secure; they should rerun security checks after each model change and provide secure APIs in the model's context.
\end{abstract}

% \begin{CCSXML}
% <ccs2012>
%  <concept>
%   <concept_id>00000000.0000000.0000000</concept_id>
%   <concept_desc>Do Not Use This Code, Generate the Correct Terms for Your Paper</concept_desc>
%   <concept_significance>500</concept_significance>
%  </concept>
%  <concept>
%   <concept_id>00000000.00000000.00000000</concept_id>
%   <concept_desc>Do Not Use This Code, Generate the Correct Terms for Your Paper</concept_desc>
%   <concept_significance>300</concept_significance>
%  </concept>
%  <concept>
%   <concept_id>00000000.00000000.00000000</concept_id>
%   <concept_desc>Do Not Use This Code, Generate the Correct Terms for Your Paper</concept_desc>
%   <concept_significance>100</concept_significance>
%  </concept>
%  <concept>
%   <concept_id>00000000.00000000.00000000</concept_id>
%   <concept_desc>Do Not Use This Code, Generate the Correct Terms for Your Paper</concept_desc>
%   <concept_significance>100</concept_significance>
%  </concept>
% </ccs2012>
% \end{CCSXML}

% \ccsdesc[500]{Do Not Use This Code~Generate the Correct Terms for Your Paper}
% \ccsdesc[300]{Do Not Use This Code~Generate the Correct Terms for Your Paper}
% \ccsdesc{Do Not Use This Code~Generate the Correct Terms for Your Paper}
% \ccsdesc[100]{Do Not Use This Code~Generate the Correct Terms for Your Paper}

% \keywords{Do, Not, Use, This, Code, Put, the, Correct, Terms, for, Your, Paper}

% \received{20 February 2007}
% \received[revised]{12 March 2009}
% \received[accepted]{5 June 2009}

\maketitle

\section{Introduction}\label{sec:int}
Large Language Models (LLMs) have become part of everyday software development. Code generation is the most studied use of LLMs in software engineering~\cite{hou2024large}, and coding assistants such as OpenAI Codex~\cite{openai2025codex} and Gemini CLI~\cite{google2025geminicli} increasingly generate and modify code on behalf of developers. To choose among the many available models, practitioners rely on leaderboards~\cite{xia2024evoeval}. For coding, these rank models mainly by how often their code passes functional tests, as on SWE-bench~\cite{jimenez2024swebench}, LiveCodeBench~\cite{jain2024livecodebench}, and HumanEval~\cite{chen2021evaluating}, and even these rankings are fragile: small variations in benchmark problem formulations can largely affect the final rank~\cite{xia2024evoeval}.
The consideration of security properties of the generated code rarely enters such rankings.
% 
% Passing functional tests, however, does not make code secure.
Early studies found that about 40\% of the programs generated by GitHub Copilot were vulnerable~\cite{pearce2025asleep} and that only 5 of 21 programs written by ChatGPT were secure without further prompting~\cite{khoury2023secure}. 
Critically, such code flaws do not remain isolated, but they reach public GitHub projects~\cite{fu2025security}, code generated by ChatGPT is merged into open-source projects~\cite{siddiq2024quality}, and developers who use an AI assistant write less secure code while being more confident in its security~\cite{perry2023users}. Novices in particular overestimate the reliability of AI output and spend less time verifying it~\cite{sergeyuk2026human}.

Although the security properties of generated code are paramount interest, current related studies evaluate the models available at one point in time~\cite{tihanyi2025secure,kharma2026security,toth2024llms}, and security benchmarks rank models only once~\cite{siddiq2022securityeval,tony2023llmseceval,hajipour2024codelmsec,vero2025baxbench,peng2025cweval}. Models, however, change with every new version~\cite{chen2023chatgpt,liu2026robustness}, and a property that changes over time can only be understood by following it across versions~\cite{kemerer1999empirical}, as a recent review of AI assistance also calls for~\cite{sergeyuk2026human}. A snapshot cannot tell whether a newer version that writes better code also writes more secure code, or whether an upgrade reintroduces weaknesses that its predecessor had avoided.
Currently, the community lacks a robust methodology to assess whether families of models are improving the security properties of their generated code, and if such improvements are biased toward specific issues or programming languages.

\citet{cannavale2026smarter} took a first step toward answering this question. They measure the \emph{functional\-ity-security gap}, the difference between how often a model generates functionally plausible code and how often that code also passes a set of security tests. They follow it over four versions each of GPT, Claude Sonnet, and Llama on CWEval~\cite{peng2025cweval}, whose dynamic oracles judge functionality and security on the same program. They conclude that models become ``smarter, but not safer'': the gap narrows for the two proprietary families but not for Llama. 
However, their design captures only part of how current models evolve. In particular, we identified three limitations.
First, they evaluate one tier per family, although vendors offer smaller and cheaper models next to their flagships, which developers may choose to reduce costs or to run a model locally. Second, their open-weight side is Llama alone, so the contrast between proprietary and open-weight models rests on a single family, while families such as Qwen, DeepSeek, Z.ai, Moonshot, and MiniMax have become competitive on coding benchmarks. Third, they examine programming languages only as a robustness check, through a single gap per language that averages over all weaknesses, although the weaknesses of generated code differ by language~\cite{kharma2026security}.

In this paper, we propose a new model methodology to assess how models of different nature (e.g., open vs closed, flagship vs compact) change their performance over time and across languages.
To achieve our goal, we study the functionality-security gap across 32 models from seven families, two proprietary (Google and OpenAI) and five open-weight (Qwen, DeepSeek, Z.ai, Moonshot, and MiniMax), over three successive versions (\emph{stages}) each and, where a family offers one, at both a flagship and a compact tier. With 100 samples for each of the 119 CWEval tasks per model, we analyze 380,800 programs to answer the following research questions:

\begin{enumerate}[start=1,label={$\mathbf{RQ_{\arabic*}}$},leftmargin=3em]
\item \textbf{(Gap Trajectory)} How does the functionality-security gap evolve across the stages of a model family?
\item \textbf{(Model Tiers)} How does the functionality-security gap differ between the flagship and the compact model of a family?
\item \textbf{(CWE Evolving Behavior)} Which CWE categories persist, diverge, or regress as model families evolve?
\item \textbf{(Programming Languages)} How do CWE categories differ across programming languages?
\end{enumerate}

Our results show that newer models do become safer in absolute terms: in every family, the newest stage produces more code that is both functionally plausible and secure\footnote{We consider a code as \textit{secure} if it \textit{passes} a set of targeted security tests, following the practice of CWEval~\cite{peng2025cweval} (Section~\ref{sec:bench}).} than the first. Every family also narrows the gap significantly by its newest stage for at least one number of attempts per task, but none closes it. Where functional plausibility grows almost as fast as security, the gap can persist, as for Google, whose newest stage does not narrow the gap significantly when models get a single attempt ($RQ_1$). When the tiers differ significantly, the compact model is almost always less secure, a penalty that shrinks as families evolve ($RQ_2$). Log injection (CWE-117) and HTTP response splitting (CWE-113) remain insecure in most models, other categories divide the families regardless of whether their weights are open, and the newest models regress on vulnerability categories that their predecessors had already handled ($RQ_3$). Finally, the risk of a category depends strongly on the language, and C has the highest vulnerability rate in most models ($RQ_4$).

These results refine and extend the findings of~\citet{cannavale2026smarter} in three ways. First, openness does not separate the families: every open-weight family narrows its gap significantly, while the newest Google flagship keeps a gap as wide as the one reported for Llama. Developers can therefore consider open-weight models without assuming a security penalty. Second, a cheaper tier is not a fixed security cost, since the compact penalty shrinks over stages and some compact models match their flagships (e.g., Gemini 3.7 Flash). Third, while the gap of each family evolves similarly across languages, individual CWEs do not: the same CWE carries different risks in different languages, so the best model choice can depend on the language and on the weaknesses one cares about most.

Our results carry direct implications. Vendors should report security scores next to functional ones and compare each new version with its predecessor per weakness type, since an aggregate improvement can hide a regression. Developers should not treat an upgrade or a cheaper tier as neutral for security, and they can steer models toward secure code by placing secure APIs in the prompt context, since models follow the path that the prompt or the library offers. Researchers should report results for multiple attempts per task and for more than two versions per family (trajectories), since both choices matter.
In summary, we make the following contributions:
\begin{itemize}[leftmargin=2em]
    \item A longitudinal study of the functionality-security gap across 32 LLMs from seven families over three stages, including a comparison of flagship and compact models within each family.
    \item An analysis of how CWE categories persist, diverge, or regress across stages, and how their risk varies across programming languages.
    \item A comparison with prior work showing that openness does not determine the security trajectory of a model family.
    \item Actionable implications for model vendors, developers, and researchers.
    \item A replication package with the generated code, execution results, and analysis scripts.
\end{itemize}

\section{Background \& Related Work}\label{sec:rel}
\subsection{Common Weakness Enumeration (CWE)}\label{sec:cwe}
The Common Weakness Enumeration (CWE), maintained by MITRE~\cite{cwe}, catalogs types of software and hardware weaknesses that can lead to vulnerabilities, each with a numeric identifier. Two weaknesses recur throughout our results.
\emph{Log injection} (CWE-117) occurs when external input is written to a log without neutralizing line breaks, which lets an attacker forge log entries.
\emph{HTTP response splitting} (CWE-113) occurs when CR and LF characters reach an outgoing HTTP header, which gives an attacker control over the remaining headers and the response body.
We group the further weaknesses in our results by their underlying mistake. In cross-site scripting (CWE-79), XPath injection (CWE-643), and eval injection (CWE-95), unescaped input reaches a web page, a query, or an interpreter. In path traversal (CWE-22) and server-side request forgery (CWE-918), input selects an unintended file or network destination. Improper verification of cryptographic signature (CWE-347) and deserialization of untrusted data (CWE-502) accept data whose origin or content is not checked. Uncontrolled resource consumption (CWE-400) lets crafted input exhaust computing resources. Out-of-bounds read (CWE-125) and integer overflow (CWE-190) are memory and arithmetic errors typical of C.

\subsection{Evaluating the Security of Generated Code}

Since working code is not necessarily secure, security has to be measured separately. \citet{tihanyi2025secure} formally verify 331,000 compilable C programs from nine models and find at least 62.07\% of them vulnerable, with only minor differences between models. \citet{kharma2026security} show that weaknesses differ by language, with memory management issues most frequent in C and C++, and cryptographic and XML weaknesses most frequent in Java and Python. \citet{toth2024llms} find exploitable vulnerabilities in at least 11.16\% of 2,500 PHP websites generated by GPT-4.

Security benchmarks detect weaknesses either statically, by matching the source code against known insecure patterns, or dynamically, by running the code on attack inputs. Early benchmarks are static: SecurityEval~\cite{siddiq2022securityeval}, LLMSecEval~\cite{tony2023llmseceval}, and CodeLMSec~\cite{hajipour2024codelmsec} use CodeQL, and CyberSecEval~\cite{bhatt2023purple} uses its own detector. Static detection misses insecure code that matches no rule. On code generated by GitHub Copilot, the authors of SecurityEval found the targeted weakness in 18.46\% of samples with CodeQL but in 73.84\% by manual inspection, and the detector of CyberSecEval reaches a recall of 79\% against manual labels. More importantly, these benchmarks do not check whether the code works. A model can avoid every weakness with a solution that does nothing~\cite{peng2025cweval}, and methods for secure code generation often trade functionality for security~\cite{dai2025rethinking}. Functionality and security must therefore be assessed together.
Recent benchmarks do so dynamically. BaxBench~\cite{vero2025baxbench} asks for complete backend applications and attacks those that pass its functional tests with end-to-end exploits, which succeed against around half of them. SeCodePLT~\cite{nie2026secodeplt} classifies each sample as failing, working but insecure, or working and secure across 44 risk categories. SEC-bench~\cite{lee2025secbench} builds tasks from real CVE reports and scores agents on exploiting and repairing the vulnerability. We use CWEval~\cite{peng2025cweval} that asks for standalone functions whose descriptions involve a security-relevant operation without mentioning security. Its single-function tasks isolate the model's output, which our study requires. Each task comes with a functional and a dynamic security oracle, and with two reference implementations, one passes only the functional oracle and one passes both (Section~\ref{sec:bench}).

\subsection{Longitudinal Studies of LLMs}

It is a common understanding that model behavior changes between versions. For example, \citet{chen2023chatgpt} compare the March and June 2023 versions of GPT-3.5 and GPT-4 on seven types of tasks and find both improvements and declines, and \citet{liu2026robustness} find that updates of GPT, Llama, and Qwen do not consistently improve adversarial robustness and that larger models are not necessarily more robust. However, both studies examine other properties, such as task performance and adversarial robustness, and neither evaluates the security of generated code.
\citet{elgedawy2025ocassionally} collect 450 samples from nine platforms of the GPT, Gemini, and DeepSeek families for nine online shopping tasks, in three versions between November 2023 and March 2025, with and without security instructions. They judge security with CodeQL and a manual review focused on the OWASP Top 10. In the final phase, every platform produced functional code for every task after revision requests, while the vulnerability counts did not decrease consistently. The study compares platforms, whose components around the model are part of what is measured, so a change between versions cannot be attributed to the model itself, and judges security only by static analysis. In contrast, we query the models directly and run each program against an attack input.

Closest to our work, \citet{cannavale2026smarter} follow 12 models from GPT, Claude Sonnet, and Llama, four versions per family, on CWEval. They introduce the functionality-security gap and the Conditional Vulnerability Rate (CVR), fit a GEE to the gap across versions, and classify vulnerability categories into persistent blind spots, evolutionary divergences, and regression anomalies.
While \citet{cannavale2026smarter} show promising results, they still have shortcomings that we address in our work. 
While we adopt their metrics, statistical approach, and patterns, we
demonstrate that a larger and more in-depth evaluation allows one to observe different facts that were previously ignored.
First, they evaluate one tier per family, the Sonnet tier for Claude, so the effect of choosing a smaller or cheaper model remains open, although such models reduce costs and, when open-weight, can run on local hardware. In contrast, we evaluate a flagship and a compact tier. 
Second, their open-weight side is Llama alone, while families such as Qwen, DeepSeek, Z.ai, Moonshot, and MiniMax have become competitive on coding benchmarks such as SWE-bench~\cite{jimenez2024swebench}, LiveCodeBench~\cite{jain2024livecodebench}, and HumanEval~\cite{chen2021evaluating}. We evaluate these five to further investigate the \emph{functionality-security gap} for those modern and popular model families.
Third, they draw 10 samples per task and report $k=10$, with $k=1$ as a check, providing a comparably low statistical power. We present more reliable results by drawing 100 samples and reporting $k=1$ and $k=50$.
Finally, they examine languages only as a robustness check. They compute a single score that averages all CWEs per language, while we are the first to report factual biases toward particular CWEs and programming languages.
In short, our methodology allows observing new insights that highlight the limitations of modern LLMs and direct research on current blind spots.

\section{Study Methodology}\label{sec:eva}
We evaluate the \emph{functionality-security gap}~\cite{cannavale2026smarter}, the difference between how often the code a model generates is functionally plausible and how often it is also secure.
We study how this gap changes as a family evolves across successive versions and between tiers, and which CWE categories and programming languages concentrate the resulting vulnerabilities. To this end, we answer four research questions as introduced in the introduction (Section~\ref{sec:int}):
$\mathbf{RQ_1}$ \textbf{(Gap Trajectory)} asks whether newer versions of a family become more secure, or only more capable while the gap remains. $\mathbf{RQ_2}$ \textbf{(Model Tiers)} asks whether a compact model is less secure than its flagship, and whether any difference only reflects its lower capability. $\mathbf{RQ_3}$ \textbf{(CWE Evolving Behavior)} asks which CWE categories concentrate insecurity and whether each one persists, diverges between families, or regresses as a family evolves. $\mathbf{RQ_4}$ \textbf{(Programming Languages)} asks whether the risk of a category depends on the programming language.

\subsection{Models Selected}\label{sec:models}
We selected models from seven families, two proprietary (Google and OpenAI) and five open-weight (Alibaba's Qwen, DeepSeek, Z.ai, Moonshot AI, and MiniMax). Within each family, we order models into three stages of successive public versions, which form the trajectory of the family. A tier separates the flagship model of a stage from its compact sibling, a smaller checkpoint of disclosed size for the open-weight families, or a cheaper API tier for the proprietary ones, whose sizes are not disclosed. 
To be able to observe the trajectory of models over time, we constituted a large dataset of 32 models (\autoref{tab:models}), grouped by family and tier and ordered by stage, and indicate which ones reason before answering.
Both proprietary families contribute a flagship and a compact model at each stage, 12 models in total. Their newest models, GPT-5.6 Sol and GPT-5.6 Luna for OpenAI and Gemini 3.1 Pro and Gemini 3.7 Flash for Google, power the coding agents of their vendors, Codex and Gemini CLI. As on the open-weight side, our selection thus covers models that developers use for coding in practice.
For the open-weight families, we selected three successive stages at the flagship tier, together with every compact sibling available. These families have shown increasingly competitive results on coding benchmarks such as SWE-bench~\cite{jimenez2024swebench}, LiveCodeBench~\cite{jain2024livecodebench}, and HumanEval~\cite{chen2021evaluating}, and each flagship was among the strongest open-weight models on these benchmarks when it appeared. Qwen offers a compact model at each of its three stages, Z.ai one at its middle stage (GLM 4.7 Flash, one month after GLM 4.7), and DeepSeek one at its newest stage (DeepSeek V4 Flash, published together with DeepSeek V4 Pro). Three checkpoints are specialized for code, Qwen 3 Coder at both tiers and Kimi K2.7 Code, which we refer to as Kimi K2.7 for simplicity. We treat them as regular stages of their family's trajectory.

\begin{table*}[t]
\centering
\small
\setlength{\tabcolsep}{4pt}
\caption{
The 32 selected models, 12 proprietary and 20 open-weight, with their tier, stage, reasoning mode, and snapshot date.
}
\label{tab:models}
\begin{tabular}{l @{\hspace{10pt}} l c p{5.4cm} @{\hspace{4pt}} c @{\hspace{4pt}} l}
\toprule
Family & Tier & Stage & Model & Reasoning & Snapshot \\
\midrule
\multirow{6}{*}{\textbf{Google}} & Flagship & 1 & Gemini 1.5 Pro~\cite{gemini15pro_api} & $\times$ & 2024-02-15 \\
 & Flagship & 2 & Gemini 2.5 Pro~\cite{gemini25pro_api} & \checkmark & 2025-03-25 \\
 & Flagship & 3 & Gemini 3.1 Pro~\cite{gemini31pro_api} & \checkmark & 2026-02-19 \\
 & Compact & 1 & Gemini 1.5 Flash~\cite{gemini15flash_api} & $\times$ & 2024-05-14 \\
 & Compact & 2 & Gemini 2.5 Flash~\cite{gemini25flash_api} & \checkmark & 2025-04-17 \\
 & Compact & 3 & Gemini 3.7 Flash~\cite{gemini37flash_api} & \checkmark & 2026-08-13 \\
\midrule
\multirow{6}{*}{\textbf{OpenAI}} & Flagship & 1 & GPT-4o~\cite{gpt4o_api} & $\times$ & 2024-05-13 \\
 & Flagship & 2 & GPT-5~\cite{gpt5_api} & \checkmark & 2025-08-07 \\
 & Flagship & 3 & GPT-5.6 Sol~\cite{gpt56sol_api} & \checkmark & 2026-07-09 \\
 & Compact & 1 & GPT-4o mini~\cite{gpt4omini_api} & $\times$ & 2024-07-18 \\
 & Compact & 2 & GPT-5 mini~\cite{gpt5mini_api} & \checkmark & 2025-08-07 \\
 & Compact & 3 & GPT-5.6 Luna~\cite{gpt56luna_api} & \checkmark & 2026-07-09 \\
\midrule
\multirow{6}{*}{\textbf{Qwen}} & Flagship & 1 & Qwen 3 (235B A22B)~\cite{qwen3235b_hf} & \checkmark & 2025-04-27 \\
 & Flagship$^{\dagger}$ & 2 & Qwen 3 Coder (480B A35B)~\cite{qwen3coder480b_hf} & $\times$ & 2025-07-22 \\
 & Flagship & 3 & Qwen 3.5 (397B A17B)~\cite{qwen35397b_hf} & \checkmark & 2026-02-16 \\
 & Compact & 1 & Qwen 3 (30B A3B)~\cite{qwen330b_hf} & \checkmark & 2025-04-27 \\
 & Compact$^{\dagger}$ & 2 & Qwen 3 Coder (30B A3B)~\cite{qwen3coder30b_hf} & $\times$ & 2025-07-31 \\
 & Compact & 3 & Qwen 3.5 (27B)~\cite{qwen3527b_hf} & \checkmark & 2026-02-24 \\
\midrule
\multirow{4}{*}{\textbf{DeepSeek}} & Flagship & 1 & DeepSeek V3 (671B)~\cite{deepseekv3_hf} & $\times$ & 2024-12-25 \\
 & Flagship & 2 & DeepSeek V3.2 (671B)~\cite{deepseekv32_hf} & $\times$ & 2025-12-01 \\
 & Flagship & 3 & DeepSeek V4 Pro (1.6T)~\cite{deepseekv4pro_hf} & $\times$ & 2026-04-22 \\
 & Compact & 3 & DeepSeek V4 Flash (284B)~\cite{deepseekv4flash_hf} & $\times$ & 2026-04-22 \\
\midrule
\multirow{4}{*}{\textbf{Z.ai}} & Flagship & 1 & GLM 4.5 (355B)~\cite{glm45_hf} & \checkmark & 2025-07-20 \\
 & Flagship & 2 & GLM 4.7 (355B)~\cite{glm47_hf} & \checkmark & 2025-12-22 \\
 & Flagship & 3 & GLM 5.2 (744B)~\cite{glm52_hf} & \checkmark & 2026-06-16 \\
 & Compact & 2 & GLM 4.7 Flash (30B)~\cite{glm47flash_hf} & \checkmark & 2026-01-19 \\
\midrule
\multirow{3}{*}{\textbf{Moonshot}} & Flagship & 1 & Kimi K2* (1T)~\cite{kimik2think_hf} & \checkmark & 2025-11-04 \\
 & Flagship & 2 & Kimi K2.5 (1T)~\cite{kimik25_hf} & \checkmark & 2026-01-01 \\
 & Flagship$^{\dagger}$ & 3 & Kimi K2.7 Code (1T)~\cite{kimik27_hf} & \checkmark & 2026-06-11 \\
\midrule
\multirow{3}{*}{\textbf{MiniMax}} & Flagship & 1 & MiniMax M2.1 (230B)~\cite{minimaxm21_hf} & \checkmark & 2025-12-20 \\
 & Flagship & 2 & MiniMax M2.5 (230B)~\cite{minimaxm25_hf} & \checkmark & 2026-02-12 \\
 & Flagship & 3 & MiniMax M3 (428B)~\cite{minimaxm3_hf} & \checkmark & 2026-06-02 \\
\bottomrule
\end{tabular}

\vspace{3pt}
{\footnotesize *Thinking variant, so that the family consists only of reasoning models. $^{\dagger}$Checkpoints specialized for code.}
\end{table*}

\subsection{Benchmark Selected}\label{sec:bench}
We use CWEval~\cite{peng2025cweval} as the task suite and test oracles. 
We chose CWEval over other benchmarks for two reasons. First, it evaluates the output of a single model. Frameworks at the repository level, such as SEC-bench~\cite{lee2025secbench}, evaluate an agent built around a model, whose additional layers prevent isolating the performance of the model itself. Second, its security oracle is dynamic. Benchmarks such as CodeLMSec~\cite{hajipour2024codelmsec} and LLMSecEval~\cite{tony2023llmseceval} rely on static analysis, which is prone to false positives and cannot show that a flagged weakness manifests at runtime. In CWEval, a sample fails the security oracle only when it behaves insecurely under the attack input, not because its code resembles a vulnerable pattern.
CWEval provides 119 tasks covering 31 CWEs in five programming languages (C, C++, Go, JavaScript, and Python). For each task, the model receives a partial source file with the function signature, a comment describing the intended behavior, and the imports the solution may use, and it completes the function body. Each task has two oracles driven by program outcomes. A functional test suite checks the intended behavior, and a security test runs the generated program against an input crafted to trigger the weakness the task probes. The two verdicts are recorded independently, so a sample can be functionally plausible but insecure, or secure but functionally implausible.
Apart from the completion budget, the batch interface described in Section~\ref{sec:data}, and three minor test fixes, we did not modify CWEval's prompts, oracles, or task set, so that our results remain comparable to prior work on the same benchmark. CWEval offers prompts with and without an instruction to write secure code, and we use only those without it, which reflect ordinary use. Passing a test suite shows that code is functionally plausible, not that it is correct~\cite{wang2026solved}, so we call code that passes the functional oracle \emph{functionally plausible}.

\subsection{Data Collection}\label{sec:data}

Four of the 12 proprietary models, those of the first stage (GPT-4o, GPT-4o mini, Gemini 1.5 Pro, and Gemini 1.5 Flash), were already evaluated by CWEval~\cite{peng2025cweval}, and we reuse its results for them. We queried the other eight through OpenRouter~\cite{openrouter}, a unified API gateway, using the batch mode of their providers to reduce costs, within an API budget of US\$600. This required adapting the generation code of CWEval.
We ran the 20 open-weight models with vLLM~\cite{kwon2023vllm} (v0.28.0) on a high performance computing cluster, one job per model on a single node with 8 GPUs and 8 CPU cores per GPU. To fit the available GPU memory, we used the official quantized checkpoint of each vendor (mostly in FP8 and in INT4 for the three Kimi checkpoints, and BF16 only where no official quantized version exists), as prior work shows that official low-bit quantization preserves functional benchmark accuracy and pass rates comparable to unquantized baselines~\cite{afrin2025quantization, vallecillos2026smaller}. Depending on the size of the weights, a job ran on NVIDIA A30 (24\,GB), H100 (80\,GB), or H200 (141\,GB) GPUs, the largest node providing about 1.1\,TB of GPU memory and 2.2\,TB of host memory.
The original completion budget of CWEval, 2{,}048 tokens, is too short for reasoning models, which spend tokens on reasoning before they answer. We raised it to 65{,}536 tokens, and to 32{,}768 tokens for Qwen 3 (235B) and Qwen 3 (30B), whose smaller context window must hold the prompt and the completion. The original budget is adequate for the four reused models, which do not reason before answering. With 100 samples for each of the 119 tasks, the 32 models produce $119 \times 100 \times 32 = 380{,}800$ programs.

\subsection{Evaluation Metrics}\label{sec:metrics}

Following the pass@$k$ protocol of \citet{chen2021evaluating}, also adopted by CWEval, we draw $n = 100$ samples per task and model. As in CWEval, the sampling temperature is fixed at 0.8 for every model except GPT-5 and GPT-5 mini, whose API accepts only its default temperature of 1. For task $i$, let $p_i$ be the number of samples that pass the functional oracle and $b_i$ the number that pass both the functional and the security oracle. On top of the unbiased pass@$k$ estimator, CWEval defines func@$k$, the probability that at least one of $k$ samples is functionally plausible, and func-sec@$k$, the probability that at least one is functionally plausible \textit{and} secure, averaged over the $T = 119$ tasks.
\begin{equation}
\text{func@}k = \mathbb{E}_{i}\left[1 - \frac{\binom{n-p_i}{k}}{\binom{n}{k}}\right], \qquad
\text{func-sec@}k = \mathbb{E}_{i}\left[1 - \frac{\binom{n-b_i}{k}}{\binom{n}{k}}\right].
\label{eq:funcfuncsec}
\end{equation}
We compute both metrics for $k \in \{1, 10, 50\}$ and report $k=1$ and $k=50$ in our results (for $k=10$ see our replication package). $k=1$ reflects the single completion that a developer typically accepts, and $k=50$ estimates whether a model can produce a plausible and secure solution at all, given enough attempts.
As in~\citet{cannavale2026smarter}, we define the \emph{functionality-security gap} $\Delta_{\text{sec}}(k)$ as the difference between the two metrics, and the Conditional Vulnerability Rate (CVR) as the share of insecure samples among the functionally plausible ones, computed over a group of tasks, such as one CWE or one language, where $p$ and $b$ are the sums of $p_i$ and $b_i$ over the tasks of the group.
\begin{equation}
\Delta_{\text{sec}}(k) = \text{func@}k - \text{func-sec@}k, \qquad
\text{CVR} = \frac{p - b}{p}.
\label{eq:secgap}
\end{equation}
At $k=1$, $\Delta_{\text{sec}}(k)$ is the share of samples that are functionally plausible but insecure. For larger $k$, it is the expected share of tasks whose $k$ samples include a functionally plausible solution but no secure one. We report the gap together with func@$k$ and func-sec@$k$, so that a change in the gap can be attributed to functionality or to security, and use CVR to identify the vulnerability categories in which the gap concentrates. Both metrics follow the definitions of \citet{cannavale2026smarter}.

\subsection{Data Analysis}\label{sec:analysis}

For $RQ_1$, we test whether the gap of a family changes significantly across its three stages. For each flagship model and task, we compute the gap between func@$k$ and func-sec@$k$ with Equation~\ref{eq:secgap} on that task alone, which yields 119 per-task gaps per model. We then fit one Generalized Estimating Equation (GEE)~\cite{liang1986longitudinal} per family and value of $k$, with the stage as the only predictor and the first stage as the baseline. For each later stage, the GEE estimates a coefficient $\beta$, the difference between the gap $\Delta_{\text{sec}}(k)$ of that stage and the gap of the first stage. A negative $\beta$ means that the later stage narrowed the gap, and a positive $\beta$ means that it widened it. Since all stages are evaluated on the same tasks, we cluster the observations by task with an exchangeable working correlation, so that the significance test accounts for this dependency. We use a Gaussian family with an identity link and consider $\beta$ significant at $p < 0.05$, following the approach of \citet{cannavale2026smarter}.

For $RQ_2$, we use the same approach to compare the flagship and the compact model of a stage, with one GEE per family, stage, and value of $k$. The tier replaces the stage as the predictor, with the flagship as the baseline. Here, $\beta$ is the gap of the compact model minus the gap of the flagship, so a positive $\beta$ means that the compact model has the wider gap.

For $RQ_3$, we compute the CVR per model and CWE category and examine how it changes across the stages of each family, focusing on the 16 categories with the highest mean CVR. Similar to~\citet{cannavale2026smarter}, we classify a category as a \emph{persistent blind spot} if its CVR stays high at every stage, as an \emph{evolutionary divergence} if some families reduce their CVR while others stagnate, and as a \emph{regression anomaly} if a later stage is worse than an earlier one. For persistent blind spots and regression anomalies, we manually inspect functionally plausible but insecure samples to identify the construct that makes the code insecure.

For $RQ_4$, we compute the CVR per programming language, pooled over all models for each CWE category and pooled over all categories for each model. This shows whether vulnerabilities concentrate in specific languages or spread evenly across the five languages of CWEval. To explain the largest differences, we manually inspect insecure and secure samples from the first and the newest flagship of each family.

\section{Results}\label{sec:res}
\subsection{$\mathbf{RQ_1}$: Evolution of the Functionality-Security Gap}\label{sec:rq1}

\begin{figure*}[t]
\centering
\begin{minipage}{0.495\textwidth}
\centering
\includegraphics[width=\linewidth]{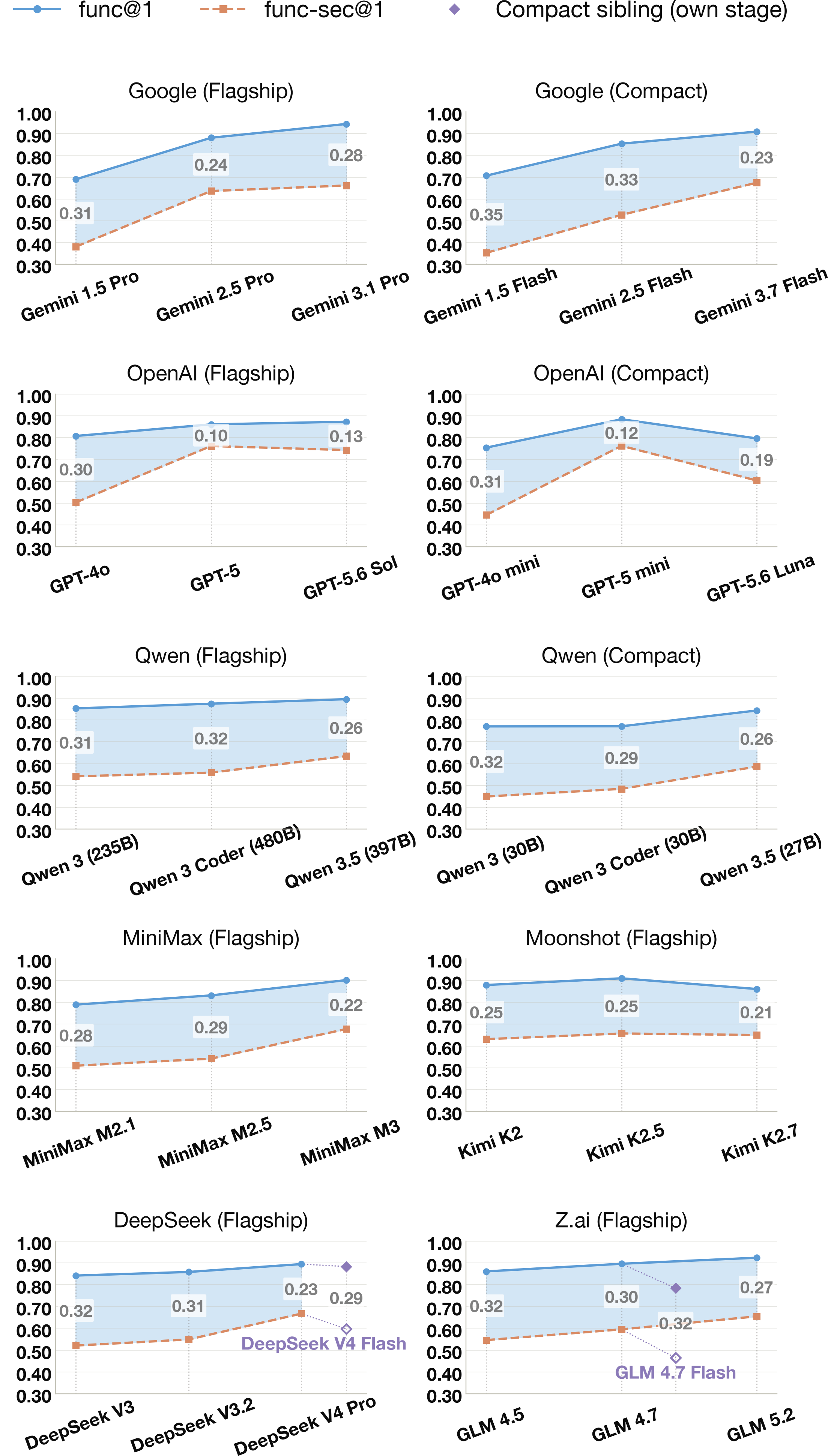}
\\[3pt]
{\footnotesize \textbf{(a) $k=1$}}
\end{minipage}%
\hfill
\vrule width 0.5pt
\hfill
\begin{minipage}{0.495\textwidth}
\centering
\includegraphics[width=\linewidth]{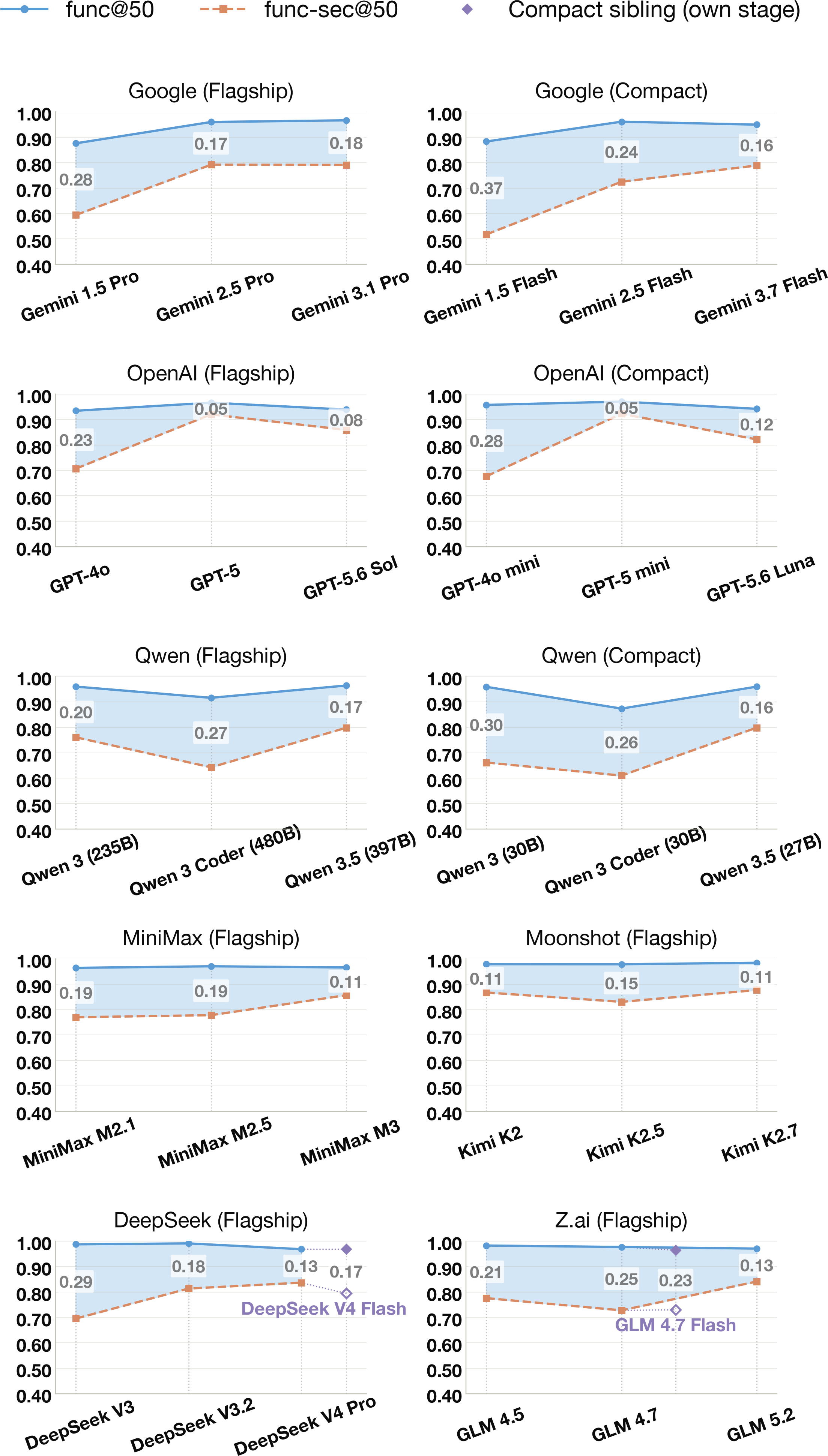}
\\[3pt]
{\footnotesize \textbf{(b) $k=50$}}
\end{minipage}
\caption{Evolution of the \emph{functionality-security gap} across the three stages of each family and tier, at $k=1$ (a) and $k=50$ (b). The shaded region between func@$k$ (solid) and func-sec@$k$ (dashed) is the gap, labeled in grey with its value at each stage. Diamonds show the compact models that exist at only one stage.}
\label{fig:gap-grid}
\end{figure*}
Across all families, a gap between func@$k$ and func-sec@$k$ persists at every stage (see \autoref{fig:gap-grid}). At the newest stage, it ranges from 0.13 (GPT-5.6 Sol) to 0.28 (Gemini 3.1 Pro) at $k=1$. At $k=50$ the gap is smaller in every family, because with more attempts more tasks obtain at least one sample that is both functionally plausible and secure, so both metrics approach their ceiling. Because functional plausibility starts high in most families and grows only moderately, the gap evolves mostly with security. Google is the exception. Its func@1 rises from 0.69 (Gemini 1.5 Pro) to 0.94 (Gemini 3.1 Pro) and its func-sec@1 by about as much, and although Google gains the most security at $k=1$, its gap barely changes. In OpenAI, by contrast, func-sec@1 rises several times as much as func@1, which gives the largest narrowing at $k=1$. At $k=50$, where Google's functional plausibility has less room to grow, its narrowing becomes clear as well. This trend indicates that a gain in security narrows the gap only when it outpaces the gain in functional plausibility.
\begin{table*}[t]
\centering
\caption{GEE results for the evolution of $\Delta_{\text{sec}}(k)$ across the flagship stages, at $k=1$ (a) and $k=50$ (b). Each family has one Gaussian GEE with the first stage as the baseline, and each row gives one of its two coefficients. A negative $\beta$ means the later stage narrowed the gap relative to the first stage. A positive $\beta$ means it widened.}
\footnotesize
\begin{minipage}{0.47\textwidth}
\centering
\textbf{(a) $k=1$}\\[3pt]
\begin{tabular}{llrrl}
\toprule
Family & Comparison & $\beta$ & $p$-value & Sig. \\
\midrule
\multirow{2}{*}{Google} & Stage 2 vs. 1 & -0.065 & 0.033 & * \\
 & Stage 3 vs. 1 & -0.027 & 0.486 & – \\
\midrule
\multirow{2}{*}{OpenAI} & Stage 2 vs. 1 & -0.205 & <0.001 & *** \\
 & Stage 3 vs. 1 & -0.175 & <0.001 & *** \\
\midrule
\multirow{2}{*}{Qwen} & Stage 2 vs. 1 & +0.005 & 0.841 & – \\
 & Stage 3 vs. 1 & -0.050 & 0.006 & * \\
\midrule
\multirow{2}{*}{DeepSeek} & Stage 2 vs. 1 & -0.011 & 0.587 & – \\
 & Stage 3 vs. 1 & -0.094 & <0.001 & *** \\
\midrule
\multirow{2}{*}{Z.ai} & Stage 2 vs. 1 & -0.014 & 0.316 & – \\
 & Stage 3 vs. 1 & -0.046 & 0.024 & * \\
\midrule
\multirow{2}{*}{Moonshot} & Stage 2 vs. 1 & +0.005 & 0.663 & – \\
 & Stage 3 vs. 1 & -0.037 & 0.015 & * \\
\midrule
\multirow{2}{*}{MiniMax} & Stage 2 vs. 1 & +0.010 & 0.401 & – \\
 & Stage 3 vs. 1 & -0.056 & 0.023 & * \\
\bottomrule
\end{tabular}
\end{minipage}
\hfill
\begin{minipage}{0.47\textwidth}
\centering
\textbf{(b) $k=50$}\\[3pt]
\begin{tabular}{llrrl}
\toprule
Family & Comparison & $\beta$ & $p$-value & Sig. \\
\midrule
\multirow{2}{*}{Google} & Stage 2 vs. 1 & -0.114 & 0.002 & * \\
 & Stage 3 vs. 1 & -0.107 & 0.006 & * \\
\midrule
\multirow{2}{*}{OpenAI} & Stage 2 vs. 1 & -0.181 & <0.001 & *** \\
 & Stage 3 vs. 1 & -0.148 & <0.001 & *** \\
\midrule
\multirow{2}{*}{Qwen} & Stage 2 vs. 1 & +0.074 & 0.005 & * \\
 & Stage 3 vs. 1 & -0.033 & 0.167 & – \\
\midrule
\multirow{2}{*}{DeepSeek} & Stage 2 vs. 1 & -0.114 & <0.001 & *** \\
 & Stage 3 vs. 1 & -0.160 & <0.001 & *** \\
\midrule
\multirow{2}{*}{Z.ai} & Stage 2 vs. 1 & +0.042 & 0.028 & * \\
 & Stage 3 vs. 1 & -0.077 & <0.001 & *** \\
\midrule
\multirow{2}{*}{Moonshot} & Stage 2 vs. 1 & +0.036 & 0.016 & * \\
 & Stage 3 vs. 1 & -0.004 & 0.753 & – \\
\midrule
\multirow{2}{*}{MiniMax} & Stage 2 vs. 1 & -0.002 & 0.921 & – \\
 & Stage 3 vs. 1 & -0.085 & 0.001 & * \\
\bottomrule
\end{tabular}
\end{minipage}
\vspace{6pt}
\par
{\footnotesize Significance codes: * $p < 0.05$; *** $p < 0.001$; -- not statistically significant.\par}
\label{tab:gee-trajectory}
\end{table*}

The results in \autoref{tab:gee-trajectory} allow two conclusions. First, every family narrows the gap significantly by its newest stage for at least one value of $k$, yet none closes it. Second, the evolution is not monotonic.
DeepSeek, Z.ai, and MiniMax narrow the gap significantly at their newest stage for both values of $k$, DeepSeek already at its middle stage for $k=50$, whereas Google does so only at $k=50$. At $k=1$, Google narrows the gap significantly only at its middle stage.
OpenAI is the only family whose two later stages both differ significantly from the first at both values of $k$, and its coefficient of $-0.205$ at the middle stage for $k=1$, a gap about 20 points narrower than at the first stage, is the largest in the table. OpenAI's newest stage does not narrow the gap further. % than its middle stage.
%, in line with the regression on CWE-125 that $RQ_3$ reports  for the newest OpenAI models.

At $k=50$, Qwen, Z.ai, and Moonshot widen the gap significantly at their middle stage, with no corresponding change at $k=1$. At this stage their func-sec@50 decreases more than their func@50, most visibly for Qwen 3 Coder (480B), the least secure Qwen flagship in $RQ_3$. Since func-sec@50 estimates the share of tasks with at least one secure sample among fifty, this widening suggests that the middle models lose security entirely on some tasks, while their single attempt remains at least as secure as before. At the newest stage all three recover. Z.ai narrows the gap significantly relative to the first stage for both values of $k$, Qwen and Moonshot only for $k=1$.

\begin{tcolorbox}[boxrule=1pt,left=2pt,right=2pt,top=2pt,bottom=2pt]
\textbf{Answer to $\mathbf{RQ_1}$.} Every family narrows the gap significantly by its newest stage (open and closed-weight models), but none closes it. Security gains reduce the gap only where they outpace gains in functional plausibility. OpenAI and DeepSeek narrow it most consistently, while Qwen, Z.ai, and Moonshot first widen it at the middle stage.
\end{tcolorbox}

\subsection{$\mathbf{RQ_2}$: Comparison Between Flagship and Compact Models}\label{sec:rq2}

While $RQ_1$ follows the gap across the stages of each family, $RQ_2$ compares the two tiers at the same stage. \autoref{fig:gap-grid} shows both tiers side by side at $k=1$ and $k=50$. Google, OpenAI, and Qwen have a compact model at every stage, while DeepSeek and Z.ai have one only at the newest and the middle stage, respectively, shown as a purple diamond. MiniMax and Moonshot have no compact model and do not enter this comparison. \autoref{tab:gee-tier} tests the difference between the tiers at each stage.

\begin{table*}[t]
\centering
\caption{GEE results for the effect of tier on $\Delta_{\text{sec}}(k)$ at each stage, at $k=1$ (a) and $k=50$ (b). Each row is a Gaussian GEE that compares the compact model of a family with its flagship at that stage, one row per available compact model. A negative $\beta$ means the compact model has the narrower gap, and a positive $\beta$ means it has the wider gap.}
\footnotesize
\begin{minipage}{0.47\textwidth}
\centering
\textbf{(a) $k=1$}\\[3pt]
\begin{tabular}{llrrl}
\toprule
Family & Comparison & $\beta$ & $p$-value & Sig. \\
\midrule
\multirow{3}{*}{Google} & Stage 1 & +0.045 & 0.126 & – \\
 & Stage 2 & +0.082 & <0.001 & *** \\
 & Stage 3 & -0.048 & 0.022 & * \\
\midrule
\multirow{3}{*}{OpenAI} & Stage 1 & +0.004 & 0.835 & – \\
 & Stage 2 & +0.023 & 0.098 & – \\
 & Stage 3 & +0.063 & 0.004 & * \\
\midrule
\multirow{3}{*}{Qwen} & Stage 1 & +0.010 & 0.647 & – \\
 & Stage 2 & -0.029 & 0.162 & – \\
 & Stage 3 & -0.004 & 0.774 & – \\
\midrule
DeepSeek & Stage 3 & +0.059 & 0.017 & * \\
\midrule
Z.ai & Stage 2 & +0.019 & 0.283 & – \\
\bottomrule
\end{tabular}
\end{minipage}
\hfill
\begin{minipage}{0.47\textwidth}
\centering
\textbf{(b) $k=50$}\\[3pt]
\begin{tabular}{llrrl}
\toprule
Family & Comparison & $\beta$ & $p$-value & Sig. \\
\midrule
\multirow{3}{*}{Google} & Stage 1 & +0.084 & 0.018 & * \\
 & Stage 2 & +0.068 & 0.010 & * \\
 & Stage 3 & -0.015 & 0.553 & – \\
\midrule
\multirow{3}{*}{OpenAI} & Stage 1 & +0.053 & 0.021 & * \\
 & Stage 2 & +0.002 & 0.914 & – \\
 & Stage 3 & +0.040 & 0.012 & * \\
\midrule
\multirow{3}{*}{Qwen} & Stage 1 & +0.098 & <0.001 & *** \\
 & Stage 2 & -0.010 & 0.737 & – \\
 & Stage 3 & -0.005 & 0.829 & – \\
\midrule
DeepSeek & Stage 3 & +0.042 & 0.075 & – \\
\midrule
Z.ai & Stage 2 & -0.015 & 0.416 & – \\
\bottomrule
\end{tabular}
\end{minipage}
\vspace{6pt}
\par
{\footnotesize Significance codes: * $p < 0.05$; *** $p < 0.001$; -- not statistically significant.\par}
\label{tab:gee-tier}
\end{table*}

The statistical results highlight two aspects. First, a compact model is not always less secure than its flagship, and second, the gap for compact models also shrinks as families evolve.
The compact model Gemini 3.7 Flash reaches the same func-sec@$k$ as its flagship Gemini 3.1 Pro in \autoref{fig:gap-grid}, and its gap at $k=1$ is significantly narrower. In 13 of the 22 overall comparisons, including every Qwen comparison at $k=1$, the two tiers do not differ significantly, and the direction of these differences is mixed, with the gap of the compact model sometimes wider and sometimes narrower than that of its flagship. When tiers do differ significantly, however, the compact model has the wider gap in eight of the nine cases.
At $k=50$, the compact models of Google, OpenAI, and Qwen all have a significantly wider gap than their flagship at the first stage, but only OpenAI still shows this difference at the newest stage.
OpenAI does not follow this convergence. Its two tiers are close at the middle stage, yet GPT-5.6 Luna has a significantly wider gap than GPT-5.6 Sol at both values of $k$, because the compact model loses security from GPT-5 mini to GPT-5.6 Luna while the flagship holds steady, in line with the regression across categories that $RQ_3$ reports for this model. DeepSeek shows the same pattern at its newest stage, significantly so at $k=1$. OpenAI and DeepSeek, the two families that narrow the gap most consistently in $RQ_1$, are also the two whose compact model has the wider gap at the newest stage.

The two curves of \autoref{fig:gap-grid} also separate two kinds of compact penalty. At $k=1$, DeepSeek V4 Flash nearly matches the functional plausibility of DeepSeek V4 Pro but produces fewer secure solutions, a security penalty at similar capability. GLM 4.7 Flash has almost the same gap as GLM 4.7 only because its functional plausibility and its security are both lower by a similar amount, which makes the smaller model weaker overall without being less secure in particular.

\begin{tcolorbox}[boxrule=1pt,left=2pt,right=2pt,top=2pt,bottom=2pt]
\textbf{Answer to $\mathbf{RQ_2}$.} When the tiers differ significantly, the compact model is less secure in eight of nine cases. The penalty shrinks as families evolve but persists at the newest stage in OpenAI and DeepSeek, whose flagships improve most consistently.
\end{tcolorbox}
\subsection{$\mathbf{RQ_3}$: Persistence, Divergence, and Regression of Vulnerabilities}\label{sec:rq3}

\begin{figure*}[htbp]
    \centering
    \includegraphics[width=\linewidth]{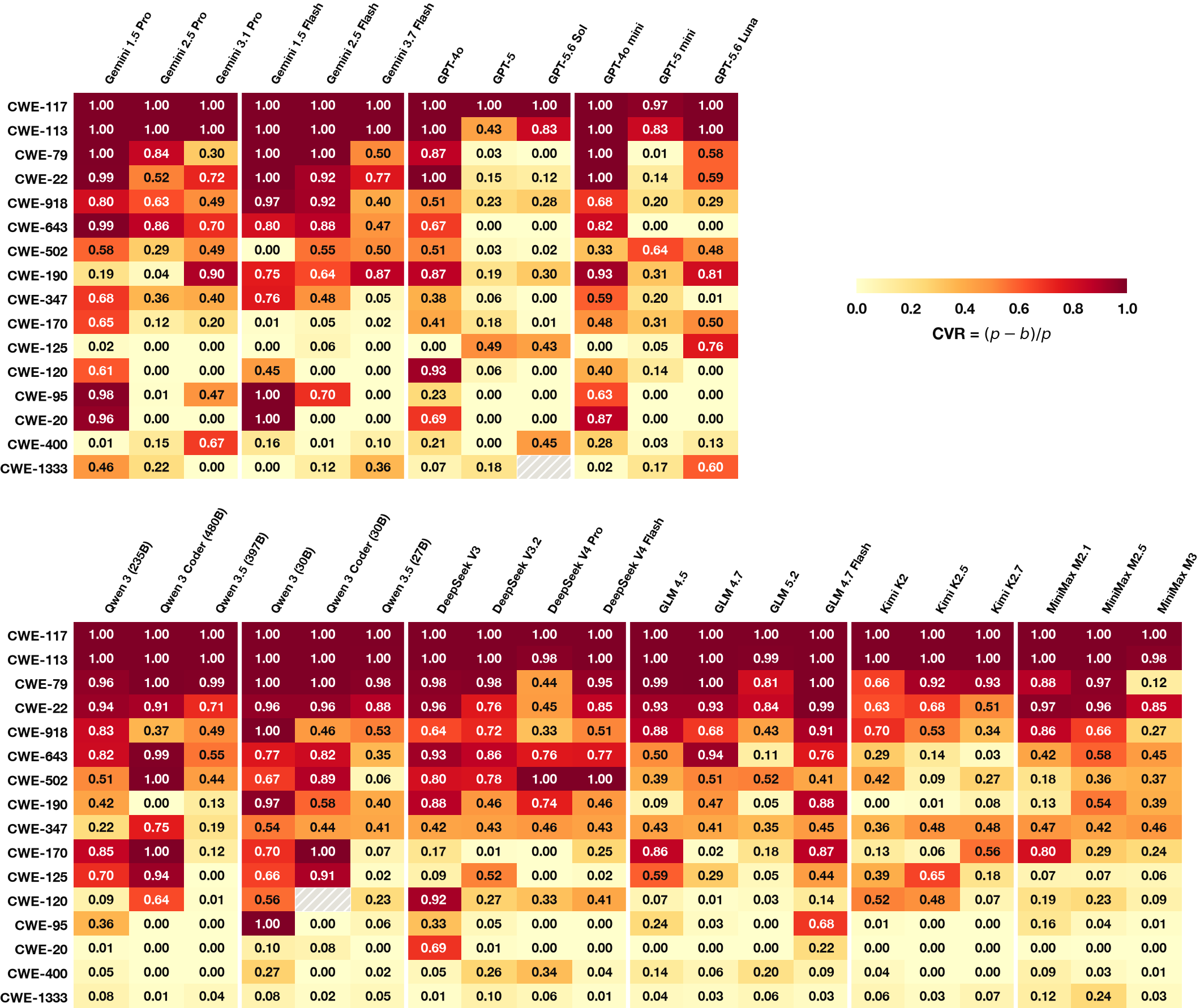}
    \caption{CVR for the 16 CWEs with the highest mean CVR, sorted from highest to lowest. Proprietary models appear in the top panel and open-weight models in the bottom panel, grouped by family and tier and ordered by stage. Hatched cells have no functionally plausible sample, so their CVR is undefined.}
    \label{fig:cvr-cwe}
\end{figure*}

While $RQ_1$ and $RQ_2$ measure the gap as a whole, $RQ_3$ asks in which vulnerability categories the gap concentrates and how each one evolves. \autoref{fig:cvr-cwe} presents the CVR of every model on the 16 CWEs with the highest mean CVR. Many categories improve as families evolve. Server-side request forgery (CWE-918) is the clearest case, ending lower than it started in all families and tiers.
The rest of this section follows the three patterns that deviate from this general improving trend.

\paragraph{Persistent Blind Spots.} Log injection (CWE-117) and HTTP response splitting (CWE-113) stay near the maximum in every family, tier, and stage. The only exception is OpenAI on CWE-113. Its middle stage lowers the rate in both tiers, and the flagship GPT-5 secures more than half of its functionally plausible samples, a gain that the newest stage partly loses. Such uniformity across independently trained families points to a blind spot shared by the field, beyond the pipeline of any single vendor. Manual inspection shows the same construction across families. The Python samples write the message into the log unchanged, as in \texttt{f"[\{timestamp\}] Received: \{msg\}"}, without removing the newline that would let an attacker forge a new entry. For CWE-113, the samples check the type of the header carefully and then assign its value without removing CRLF sequences.

\paragraph{Evolutionary Divergence.} Cross-site scripting (CWE-79) divides the open-weight families, while both proprietary families end well below where they started. The OpenAI flagship brings its rate to zero and Google cuts it substantially, and so do the DeepSeek and MiniMax flagships, the latter almost as far as OpenAI. Qwen, in contrast, stays at the maximum in both tiers, Z.ai improves only slightly, and Moonshot even gets worse. CWE-22 (path traversal) divides the families in a similar way. The OpenAI and DeepSeek flagships improve strongly, while every other newest model stays at or above one half.
XPath injection (CWE-643) separates the families even more clearly. OpenAI resolves it in both tiers, and the Moonshot and Z.ai flagships come close, while DeepSeek and the Google flagship keep most of their functionally plausible samples insecure. None of these splits follows the line between proprietary and open-weight families. CWE-347 (improper verification of cryptographic signature) is the clearest case where it does. All proprietary trajectories improve on it, the compact models most of all, whereas no open-weight trajectory shows a comparable gain.

\paragraph{Regression Anomalies.} CWE-400 (uncontrolled resource consumption) regresses in the flagships of Google, OpenAI, and DeepSeek, and manual inspection shows one shared cause. The earlier models solve the task with a plain text search, whereas the newest ones treat the caller's search pattern as a regular expression and match it against the caller's own text, so an attacker controls both sides of the match.
The newest proprietary models also regress on categories they had already handled well. OpenAI is the only family whose newest models are worse than its first ones on out-of-bounds read (CWE-125), rising from zero in GPT-4o and GPT-4o mini, while the open-weight families that spike on this category at their middle stage return below their starting point. On integer overflow (CWE-190), Gemini 3.1 Pro rises from almost zero in Gemini 2.5 Pro to the highest rate of any newest model, and the three highest rates among the newest models all belong to proprietary ones. GPT-5.6 Luna is less secure than GPT-5 mini on most of the 16 categories, at least four of them by a wide margin, and Gemini 3.1 Pro brings back eval injection (CWE-95), which Gemini 2.5 Pro had almost eliminated.
Other regressions are transient. Qwen's middle stage, Qwen 3 Coder (480B), a model specialized for code, is clearly its least secure flagship on six of the 16 categories, and the newest stage recovers on all six. Deserialization of untrusted data (CWE-502) is one of these six, but in other families the regression persists. The newest DeepSeek models of both tiers are insecure on almost every functionally plausible sample, the compact tier of Google starts fully secure and never returns to it, and Z.ai and MiniMax regress to a lesser degree.

\begin{tcolorbox}[boxrule=1pt,left=2pt,right=2pt,top=2pt,bottom=2pt]
\textbf{Answer to $\mathbf{RQ_3}$.} CWE-117 and CWE-113 remain near the maximum in almost every model, a blind spot shared across the field. CWE-79 and CWE-643 divide the families regardless of openness, whereas CWE-347 improves in all proprietary families with no comparable gain in the open-weight ones. Regressions affect CWE-400, where the newest flagships introduce regular expressions, and CWE-125 and CWE-190 in the newest proprietary models, so a newer version can lose security that its predecessor already had.
\end{tcolorbox}

\subsection{$\mathbf{RQ_4}$: Vulnerability Categories Across Programming Languages}\label{sec:rq4}

\begin{figure*}[htbp]
\centering
\begin{minipage}{0.485\textwidth}
\centering
\includegraphics[width=0.93\linewidth]{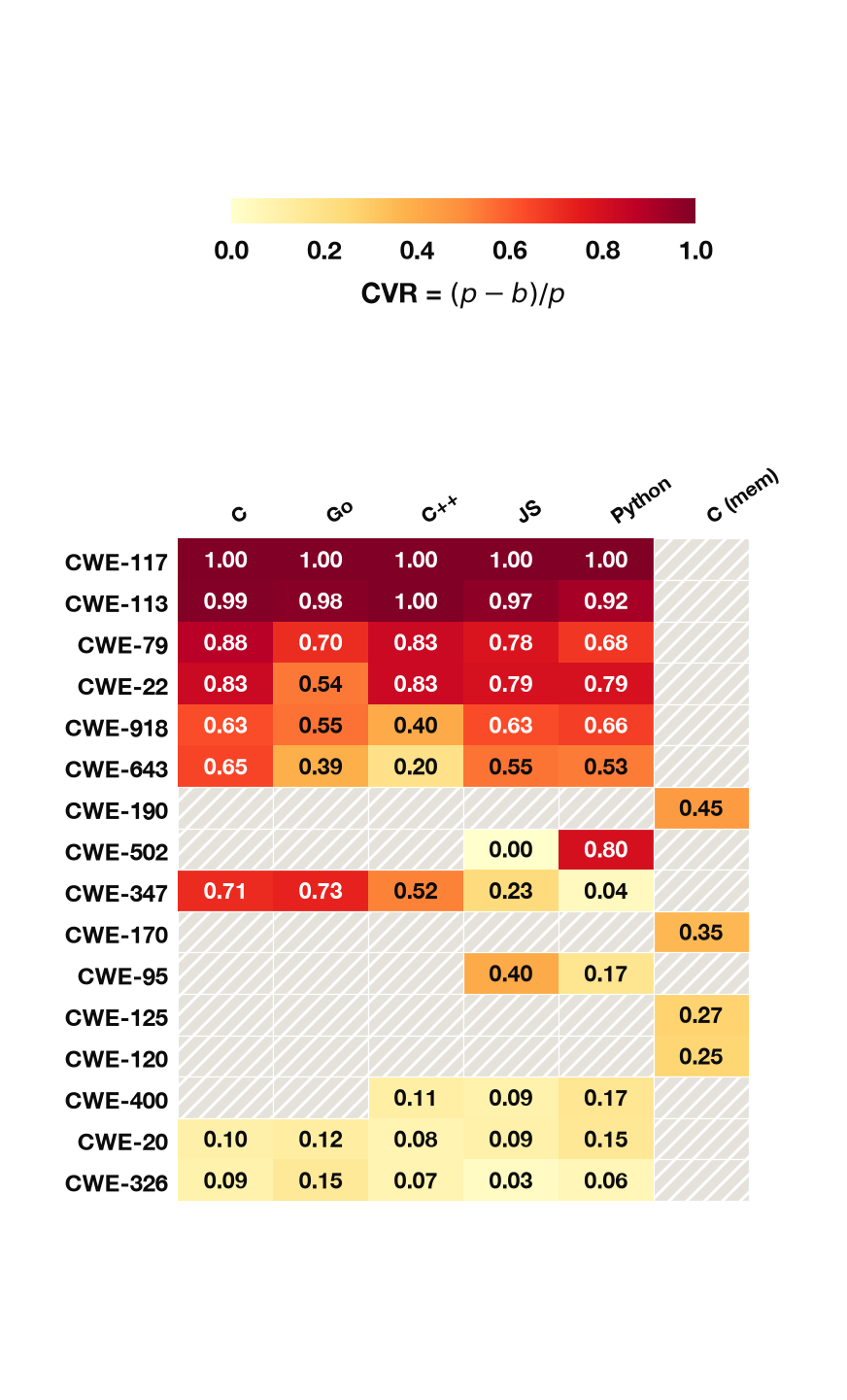}
\\[3pt]
{\footnotesize \textbf{(a) CWE $\times$ language}}
\end{minipage}%
\hfill
\vrule width 0.5pt
\hfill
\begin{minipage}{0.485\textwidth}
\centering
\includegraphics[width=0.93\linewidth]{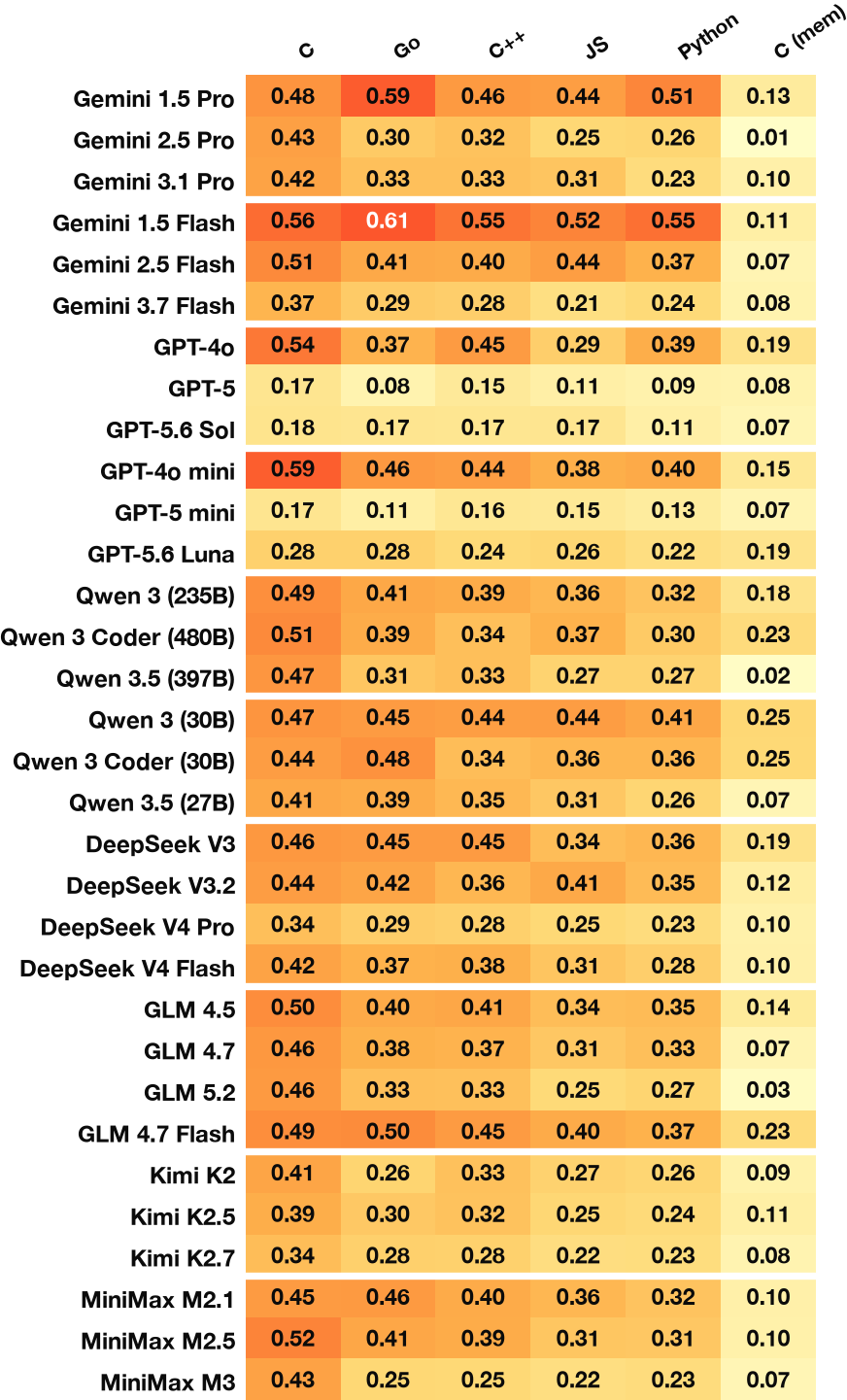}
\\[3pt]
{\footnotesize \textbf{(b) Model $\times$ language}}
\end{minipage}
\caption{CVR by programming language. (a) Pooled over all 32 models for the 16 CWEs with the highest CVR, with hatched cells where CWEval has no task. (b) Per model, grouped by family and tier.} % Both panels share the color scale in (a).
\label{fig:cvr-lang}
\end{figure*}

While $RQ_3$ traces each CWE category across the stages of a family, $RQ_4$ asks whether a category carries the same risk in every programming language. \autoref{fig:cvr-lang} (a) pools all 32 models and breaks CVR down by CWE category, and \autoref{fig:cvr-lang} (b) keeps each model separate and combines all categories of a language into one rate. We call C, Go, C++, JavaScript, and Python the core languages. The tasks dedicated to memory safety, labeled C~(mem), share no category with them and have the lowest rate in every model, so we do not discuss them further.
C has the highest average CVR, followed by Go and C++, and then by JavaScript and Python.
C is the most vulnerable core language in 27 of the 32 models, and it stays first when the comparison is restricted to the 15 of the 31 CWEs in CWEval that every core language covers.

\paragraph{Categories Shared by All Core Languages.} CWE-117 and CWE-113 stay near the maximum in every language, but the other shared categories depend strongly on the language. CWE-347 has the widest spread, with a CVR of 0.73 in Go and 0.71 in C against 0.04 in Python.
Manual samples inspection from the first and the newest flagship of each family shows that, for CWE-347, the insecure C, Go, and JavaScript samples verify the signature with the key but accept any HMAC algorithm. The Python library refuses to decode a token without an explicit algorithm, so almost every Python sample fixes \texttt{algorithms=["HS256"]}. CWE-22 is clearly lower in Go. For archive extraction, C and C++ samples reproduce the extraction example of the prompt, which has no path check, and Python samples call \texttt{extractall} without a filter, while Go samples, with neither, mostly write their own containment check. CWE-643 is lowest in C++, where most samples avoid building the XPath from the input, the construct behind almost all insecure samples.

\paragraph{Categories Limited to Some Languages.} CWE-502 and CWE-95 are tested only in Python and JavaScript, and both split the two languages. CWE-502 reaches 0.80 in Python and 0.00 in JavaScript. In the inspected models, Python samples mostly reuse the unsafe \texttt{Loader} of PyYAML that the prompt imports, and every sample that passes it to \texttt{yaml.load} is insecure, while the one sample in five that calls \texttt{yaml.safe\_load} instead is always secure. JavaScript samples use a YAML loader that is safe by default. CWE-95 reverses the order. Both languages fail by evaluating the raw input, but Python samples usually parse the expression through its syntax tree, while JavaScript, with no expression parser in its standard library, mostly relies on \texttt{eval} or \texttt{new Function}.

\paragraph{Why C Leads.} C combines a high starting point with the smallest improvement. In Google, Qwen, Z.ai, and MiniMax, C shows the smallest decrease of the five core languages between the first and the newest stage, and its distance from the other languages grows accordingly. OpenAI is the opposite case, with C improving the most and its newest models having the smallest difference between languages of any newest stage. The resistance of C concentrates on CWE-22 and CWE-79. On CWE-79, the Flash models of Google bring the rate close to zero in Go and Python, while every functionally plausible C sample of Gemini 3.7 Flash remains insecure.
Inspection of the insecure CWE-79 samples finds no escaping, as in \texttt{strcat(message, username)}. Escaping is least frequent in C, where the model must write its own function, whereas Python and Go need a single call.

\begin{tcolorbox}[boxrule=1pt,left=2pt,right=2pt,top=2pt,bottom=2pt]
\textbf{Answer to $\mathbf{RQ_4}$.} The risk of a CWE category depends strongly on the language. CWE-347 is frequent in Go and C but rare in Python, whose library enforces a fixed algorithm, and CWE-502 is frequent in Python, where models reuse the unsafe loader of the prompt. C has the highest CVR, also on the categories that all core languages share, since it barely improves on CWE-22 and CWE-79 in most families. Models therefore follow the shortest path that the language or the prompt offers, whether that path is secure or not.
\end{tcolorbox}

\section{Discussion}\label{sec:dis}
Functional plausibility is high from the first stage in our results, yet no family closes the functionality-security gap. Technical debt can harm the security of software~\cite{recupito2024technical}, and security debt has been found to accumulate quickly in LLM-based software~\cite{qi2024fine}. Generated code adds to this debt whenever it passes its functional tests but remains exploitable, and in our results this affects weaknesses of high practical relevance. CWE-79, first in the CWE Top 25~\cite{cwe_top25}, stays at the maximum in both Qwen tiers, CWE-22, sixth, stays at or above one half in every newest model except GPT-5.6 Sol and DeepSeek V4 Pro, and CWE-502, fifteenth, regresses in several families. Security therefore has to be evaluated alongside functionality instead of being assumed from it.

\subsection{Implications for LLM Vendors}

\paragraph{Security Benchmarks for Every New Version.} Vendors present a new model mainly through its functional results. Each open-weight flagship we evaluate was among the strongest open-weight models on SWE-bench~\cite{jimenez2024swebench}, LiveCodeBench~\cite{jain2024livecodebench}, or HumanEval~\cite{chen2021evaluating} when it appeared, yet none closes the gap. Our results show why a functional score cannot stand in for security. Gemini 3.1 Pro, for instance, is both more capable and more secure than Gemini 1.5 Pro, yet its gap barely changes ($RQ_1$). \citet{elgedawy2025ocassionally} observe the same across three observation phases, where the platforms they study reach full functionality while their vulnerability counts do not follow. Vendors should report func-sec@$k$ and the gap next to the functional scores of every new version.

\paragraph{Regression Checks per Weakness Type.} A narrowing gap does not mean that every weakness improves with it. OpenAI narrows the gap most consistently, yet its newest models are the only ones worse than their first on CWE-125, eighth in the CWE Top 25. DeepSeek, which also narrows its gap consistently, regresses on CWE-502 and CWE-400. An aggregate score hides such losses, and a family that improves on average can still become less secure on an individual weakness. Before publishing, vendors should compare new models with their predecessors per CWE (see \autoref{fig:cvr-cwe}).

\paragraph{Targeted Training for Persistent Blind Spots.} CWE-117 and CWE-113 remain insecure in almost every family, stage, and tier we evaluate. This is not a limit of capability. GPT-5 secures more than half of its functionally plausible samples on CWE-113, and the secure version differs from the insecure one by a single step, removing line breaks from the input. Security instruction tuning, as in SafeCoder~\cite{he2024instructiontuningsecurecode}, raises the share of secure code for the weakness types in its training data while preserving functionality. Vendors can add CWE-117 and CWE-113 to such data, which targets the two categories that no family resolves on its own as it evolves.

\subsection{Implications for Developers}

\paragraph{Switching Model Versions.} Over a full trajectory, upgrading pays off. Every family narrows the gap significantly by its newest stage, and func-sec@1 of every flagship is higher at the newest stage than at the first. A single step, however, can go the other way. Qwen, Z.ai, and Moonshot widen the gap at their middle stage, and GPT-5.6 Luna is less secure than GPT-5 mini on most of the 16 categories, and a developer who upgraded to these models lost security. Prior work observes the same for other properties, where a newer version of a model improves on some of them and gets worse on others~\cite{chen2023chatgpt,liu2026robustness}. Developers should therefore not treat an upgrade as a security improvement by default, and should rerun their security checks whenever they switch versions.

\paragraph{Choosing a Compact Model.} Choosing the compact model of a family is not only a decision about cost and speed. When the tiers differ significantly, the compact model almost always has the wider gap, and its functional behavior may give no hint of it, as DeepSeek V4 Flash shows ($RQ_2$). This penalty shrinks as families evolve, and Gemini 3.7 Flash even has a narrower gap than Gemini 3.1 Pro at $k=1$. Developers should thus compare the security of both tiers of the version they adopt instead of inferring it from the tier.

\paragraph{Matching the Model to the Code.} Which weakness a model handles worst differs between models and languages. CWE-79 divides the families regardless of openness, and C has the highest CVR in most models ($RQ_4$). Developers should choose a model by the weaknesses that matter for their code and review generated C code with particular care. For the two blind spots, no model choice helps, and code that writes logs or HTTP headers needs manual review whichever model produced it.

\paragraph{Offering a Secure Path.} Models follow the path that the language or the prompt offers ($RQ_4$). Almost every Python sample fixes the HMAC algorithm on CWE-347, because the library requires it, whereas on CWE-502 Python samples reuse the unsafe YAML loader that the prompt imports, and JavaScript samples, whose YAML loader is safe by default, leave no functionally plausible sample insecure. The context of a request thus shapes its security. Developers can turn this tendency to their advantage by placing secure APIs in that context, such as a safe loader or a validation helper, and library maintainers can make the secure call the default one, which, as the JavaScript YAML loader shows, can remove a weakness entirely.

\subsection{Implications for Researchers}

\paragraph{Evaluation Design Shapes the Conclusions.} Which families narrow their gap significantly depends on the number of attempts. Google narrows significantly by its newest stage only at $k=50$, and the widening of Qwen, Z.ai, and Moonshot at the middle stage appears at $k=50$ but not at $k=1$. The two values measure different things, whether a single sample is secure and whether any of many samples is, so studies should report both. The number of versions matters as well. A comparison of only the first and the newest version would show Qwen, Z.ai, and Moonshot narrowing and miss the widening in between, so longitudinal studies need more than two versions per family. Finally, the choice of families decides the conclusions about open-weight models, since DeepSeek and Moonshot evolve in opposite directions on the same weakness, as CWE-79 shows.

\paragraph{Preventing Security Regression Across Versions.} The regressions we observe, in the gap at the middle stage and in individual weakness types, leave open why security is lost between versions. Fine-tuning on benign data can weaken the safety alignment of an LLM unevenly across categories~\cite{qi2024fine}, and \citet{liu2026robustness} find that safety fine-tuning improves resistance to jailbreaks while degrading robustness on classification tasks. Our manual inspection points to one concrete mechanism, since the newest flagships of three families replace a plain text search with a regular expression built from the pattern of the caller (CWE-400). Whether the training of newer models favors such general constructs, and how to prevent the resulting regressions, remains to be studied.

\subsection{Relation to Prior Longitudinal Work}

\citet{cannavale2026smarter} follow GPT, Claude Sonnet, and Llama over four versions each on the same benchmark, and several of their results hold in our seven families. No family closes the gap in either study, CWE-117 and CWE-113 stay near the maximum in both, and both studies trace CWE-117 to the same omission, a log message written without neutralizing line breaks. The newest proprietary models also regress on CWE-125 in both studies. Their GPT trajectory is close to ours as well, with the gap at $k=1$ narrowing from 0.360 to 0.144 there and from 0.30 to 0.13 here. The two studies differ, however, in three respects.

First, our results refine their central finding that models become smarter but not safer. Across our families, func-sec@1 rises from the first to the newest stage in every family, so newer models do become safer in absolute terms. They become safer in proportion to their capability only when their security grows faster than their functional plausibility, which Google, with the largest gain in security at $k=1$, shows does not always happen.

Second, our results disagree with theirs on open-weight models. \citet{cannavale2026smarter} report a clean split, in which GPT and Claude narrow the gap significantly while Llama shows no significant movement, and they read CWE-22 and CWE-79 as weaknesses that proprietary models resolve and Llama does not. In our results, openness does not separate the families. Every open-weight family narrows the gap significantly by its newest stage for at least one value of $k$, and DeepSeek, significant at both stages for $k=50$, matches OpenAI there. Google, in turn, does not narrow its gap significantly at its newest stage for $k=1$, and Gemini 3.1 Pro keeps a gap of 0.28, close to the 0.296 they report for Llama 4 Scout. The same holds per weakness. On CWE-79, Qwen stays at the maximum as Llama does, while DeepSeek and MiniMax reduce their rate sharply, and on CWE-22 both Google models of the newest stage keep most plausible samples insecure, while DeepSeek V4 Pro improves strongly. For us, only CWE-347 separates proprietary from open-weight families.

We attribute this difference mainly to the choice of families. A single lineage cannot represent the open-weight field, and \citet{cannavale2026smarter} themselves name DeepSeek among the models to which their findings may not generalize. The path of Llama also resembles the one we observe for Qwen, Z.ai, and Moonshot. Its security drops at Llama 3.3 before recovering, a widening at an intermediate version like the one that becomes significant in our results only at $k=50$, so the stagnation they report may hide a similar path.

Finally, the two studies differ on programming languages. \citet{cannavale2026smarter} find the trajectories of the gap uniform across the five languages and conclude that language is not a primary discriminator. At the level of individual weaknesses, our results show the opposite. The same CWE carries very different risk in different languages, as CWE-347 in Go and Python shows, and C leads in most models, with a lead that grows in four families. A uniform trajectory of the gap per language can thus hide strong differences between languages at the level of individual weaknesses. Taken together, the two studies agree on the persistence of the gap and its blind spots. Moreover, our study shows that newer models do become safer in absolute terms, that openness does not decide how a family evolves, that compact models can be a viable alternative, and that the risk of a weakness depends on the programming language.

\section{Threats to Validity}\label{sec:thr}
\paragraph{Internal Validity}
\textbf{(Data Contamination)}
CWEval predates most models we evaluate, so its tasks may have entered their training data. This could inflate the improvement across stages, but it cannot explain the persistent weaknesses and regressions: even the newest models remain near the maximum CVR on CWE-113 and CWE-117 and regress on categories their predecessors had already handled ($RQ_3$).
\textbf{(Prompt Engineering)}
We did not alter the prompts of CWEval, which keeps our results comparable with prior work and measures the default behavior of the models, the one developers get when they leave security implicit~\cite{elahi2011security}. Security-aware prompts may yield a smaller gap, which we leave for future work.
\textbf{(Model Configurations)}
We used consistent parameter configurations throughout all experiments when possible. We drew 100 samples per task and model at the CWEval temperature of $T=0.8$, except for GPT-5 and GPT-5 mini, whose API only accepts $T=1$. 
For the four proprietary models of the first stage, we reuse the CWEval results (same prompts, oracles, and temperature), since Google no longer serves Gemini 1.5 Pro and Flash. Open-weight models ran in the official quantized precision that vendors publish for deployment.
\textbf{(Manual Inspection)}
The analyses for $RQ_3$ and $RQ_4$ are fully automated. Manual inspection by two co-authors served only to explain the identified patterns, and we report a cause only when it recurs across the inspected samples.
\textbf{(Differences in Model Stages)}
Stages within a family also differ in size, code specialization, and reasoning. We keep these differences as part of how vendors evolve their models, and use the thinking variant of Kimi K2 so that all Moonshot stages reason. The observed gap does not simply follow size: at $k=50$, it narrows significantly between DeepSeek V3 and V3.2, with equal size. Since the intervals between versions are irregular (5 to 27 months per trajectory), we describe the change per version, which matches how developers adopt new models.

\paragraph{External Validity}
\textbf{(Model Families)}
Our findings may not generalize to other model families, CWEs, or programming languages. We omit Claude and Llama because prior work already covers their trajectories on CWEval~\cite{cannavale2026smarter}, and we relate our findings to theirs in our discussion.
\textbf{(Model Integrations)}
CWEval tasks ask for a single function tested in isolation, whereas real projects embed generated code in larger codebases. Such tasks are required to isolate the output of a model (Section~\ref{sec:bench}); whether a larger context helps avoid the observed weaknesses remains future work.

\paragraph{Conclusion Validity}
\textbf{(Number of Tasks)}
Several CWEs rest on only one or two tasks, so per-CWE results may reflect task-specific properties. We therefore treat them as observations on the benchmark tasks, not as general claims about the CWE. To reduce the risk of spurious effects, we manually inspected the regressions and found that they share a concrete pattern across models, such as building a regular expression from the pattern of the caller for CWE-400. Cells without any functionally plausible sample are hatched in our figures, and none of our claims rests on a cell with few such samples.
\textbf{(Aggregate Analysis)}
Since each task appears at every stage and tier, we fit a GEE clustered by task. With 119 clusters, its coefficient estimates and cluster-robust standard errors remain valid even if the assumed within-task correlation structure is incorrect~\cite{liang1986longitudinal}.

\section{Conclusion}\label{sec:con}
In this work, we studied how the \emph{functionality-security gap} of code generated by LLMs evolves across 32 models from seven families, two proprietary and five open-weight, over three stages and two tiers. Every family narrows the gap significantly by its newest stage, but none closes it, and the gap narrows only where security grows faster than functional plausibility. Openness does not decide how a family evolves. When the tiers differ significantly, the compact model is less secure in eight of nine cases. CWE-117 and CWE-113 remain insecure in almost every model, newer versions can regress on individual weaknesses, and C has the highest CVR in most models. In future work, we plan to extend our evaluation with the secure mode of CWEval, whose prompts include an instruction to write secure code, and to compare how this instruction changes the generation of vulnerable code across stages and tiers. We also plan to study how the security of code produced by coding agents evolves across their successive versions, keeping the underlying model fixed to separate the effect of the agent from that of the model.

\section{Data Availability}
The data underlying this article are available from the corresponding author upon reasonable request.

\begin{acks}
We gratefully acknowledge the HPC resources provided by the HPC cluster Elysium of the Ruhr University Bochum, subsidized by the DFG (INST 213/1055-1).
We also gratefully acknowledge the CWEval \cite{peng2025cweval} authors for providing the data generated in their paper for the model executions studied, which enable a direct comparison with the proprietary family first stages.
\end{acks}

\bibliographystyle{ACM-Reference-Format}
\bibliography{references, models}

@misc{openrouter,
author={{OpenRouter}},
title={{OpenRouter}},
howpublished={\url{https://openrouter.ai}},
note={Accessed: September 2026},
year={2026},
}

@misc{kimik2think_hf,
author={{Moonshot AI}},
title={{Kimi-K2-Thinking}},
howpublished={\url{https://huggingface.co/moonshotai/Kimi-K2-Thinking}},
note={Published: November 2025. Accessed: September 2026},
year={2025},
month=nov,
}

@misc{kimik25_hf,
author={{Moonshot AI}},
title={{Kimi-K2.5}},
howpublished={\url{https://huggingface.co/moonshotai/Kimi-K2.5}},
note={Published: January 2026. Accessed: September 2026},
year={2026},
month=jan,
}

@misc{kimik27_hf,
author={{Moonshot AI}},
title={{Kimi-K2.7-Code}},
howpublished={\url{https://huggingface.co/moonshotai/Kimi-K2.7-Code}},
note={Published: June 2026. Accessed: September 2026},
year={2026},
month=jun,
}

@misc{minimaxm21_hf,
author={{MiniMax AI}},
title={{MiniMax-M2.1}},
howpublished={\url{https://huggingface.co/MiniMaxAI/MiniMax-M2.1}},
note={Published: December 2025. Accessed: September 2026},
year={2025},
month=dec,
}

@misc{minimaxm25_hf,
author={{MiniMax AI}},
title={{MiniMax-M2.5}},
howpublished={\url{https://huggingface.co/MiniMaxAI/MiniMax-M2.5}},
note={Published: February 2026. Accessed: September 2026},
year={2026},
month=feb,
}

@misc{minimaxm3_hf,
author={{MiniMax AI}},
title={{MiniMax-M3-MXFP8}},
howpublished={\url{https://huggingface.co/MiniMaxAI/MiniMax-M3-MXFP8}},
note={Published: June 2026. Accessed: September 2026},
year={2026},
month=jun,
}

@misc{glm45_hf,
author={{Z.ai}},
title={{GLM-4.5-FP8}},
howpublished={\url{https://huggingface.co/zai-org/GLM-4.5-FP8}},
note={Published: July 2025. Accessed: September 2026},
year={2025},
month=jul,
}

@misc{glm47_hf,
author={{Z.ai}},
title={{GLM-4.7-FP8}},
howpublished={\url{https://huggingface.co/zai-org/GLM-4.7-FP8}},
note={Published: December 2025. Accessed: September 2026},
year={2025},
month=dec,
}

@misc{glm47flash_hf,
author={{Z.ai}},
title={{GLM-4.7-Flash}},
howpublished={\url{https://huggingface.co/zai-org/GLM-4.7-Flash}},
note={Published: January 2026. Accessed: September 2026},
year={2026},
month=jan,
}

@misc{glm52_hf,
author={{Z.ai}},
title={{GLM-5.2-FP8}},
howpublished={\url{https://huggingface.co/zai-org/GLM-5.2-FP8}},
note={Published: June 2026. Accessed: September 2026},
year={2026},
month=jun,
}

@misc{gpt4o_api,
author={{OpenAI}},
title={{GPT-4o}},
howpublished={\url{https://developers.openai.com/api/docs/models/gpt-4o}},
note={Published: May 2024. Accessed: September 2026},
year={2024},
month=may,
}

@misc{gpt4omini_api,
author={{OpenAI}},
title={{GPT-4o mini}},
howpublished={\url{https://developers.openai.com/api/docs/models/gpt-4o-mini}},
note={Published: July 2024. Accessed: September 2026},
year={2024},
month=jul,
}

@misc{gpt5_api,
author={{OpenAI}},
title={{GPT-5}},
howpublished={\url{https://openai.com/index/introducing-gpt-5/}},
year={2025},
month=aug,
note={Published: August 7, 2025. Accessed: September 2026},
}

@misc{gpt5mini_api,
author={{OpenAI}},
title={{GPT-5 mini}},
howpublished={\url{https://developers.openai.com/api/docs/models/gpt-5-mini}},
year={2025},
month=aug,
note={Published: August 7, 2025. Accessed: September 2026},
}

@misc{gpt56sol_api,
author={{OpenAI}},
title={{GPT-5.6 Sol}},
howpublished={\url{https://developers.openai.com/api/docs/models/gpt-5.6-sol}},
year={2026},
month=jul,
note={Published: July 2026. Accessed: September 2026},
}

@misc{gpt56luna_api,
author={{OpenAI}},
title={{GPT-5.6 Luna}},
howpublished={\url{https://developers.openai.com/api/docs/models/gpt-5.6-luna}},
year={2026},
month=jul,
note={Published: July 2026. Accessed: September 2026},
}

@misc{deepseekv3_hf,
author={{DeepSeek AI}},
title={{DeepSeek-V3}},
howpublished={\url{https://huggingface.co/deepseek-ai/DeepSeek-V3}},
note={Published: December 2024. Accessed: September 2026},
year={2024},
month=dec,
}

@misc{deepseekv32_hf,
author={{DeepSeek AI}},
title={{DeepSeek-V3.2}},
howpublished={\url{https://huggingface.co/deepseek-ai/DeepSeek-V3.2}},
note={Published: December 2025. Accessed: September 2026},
year={2025},
month=dec,
}

@misc{deepseekv4pro_hf,
author={{DeepSeek AI}},
title={{DeepSeek-V4-Pro}},
howpublished={\url{https://huggingface.co/deepseek-ai/DeepSeek-V4-Pro}},
note={Published: April 2026. Accessed: September 2026},
year={2026},
month=apr,
}

@misc{deepseekv4flash_hf,
author={{DeepSeek AI}},
title={{DeepSeek-V4-Flash}},
howpublished={\url{https://huggingface.co/deepseek-ai/DeepSeek-V4-Flash}},
note={Published: April 2026. Accessed: September 2026},
year={2026},
month=apr,
}

@misc{qwen3235b_hf,
author={{Qwen Team}},
title={{Qwen3-235B-A22B-FP8}},
howpublished={\url{https://huggingface.co/Qwen/Qwen3-235B-A22B-FP8}},
note={Published: April 2025. Accessed: September 2026},
year={2025},
month=apr,
}

@misc{qwen330b_hf,
author={{Qwen Team}},
title={{Qwen3-30B-A3B-FP8}},
howpublished={\url{https://huggingface.co/Qwen/Qwen3-30B-A3B-FP8}},
note={Published: April 2025. Accessed: September 2026},
year={2025},
month=apr,
}

@misc{qwen3coder480b_hf,
author={{Qwen Team}},
title={{Qwen3-Coder-480B-A35B-Instruct-FP8}},
howpublished={\url{https://huggingface.co/Qwen/Qwen3-Coder-480B-A35B-Instruct-FP8}},
note={Published: July 2025. Accessed: September 2026},
year={2025},
month=jul,
}

@misc{qwen3coder30b_hf,
author={{Qwen Team}},
title={{Qwen3-Coder-30B-A3B-Instruct-FP8}},
howpublished={\url{https://huggingface.co/Qwen/Qwen3-Coder-30B-A3B-Instruct-FP8}},
note={Published: July 2025. Accessed: September 2026},
year={2025},
month=jul,
}

@misc{qwen35397b_hf,
author={{Qwen Team}},
title={{Qwen3.5-397B-A17B-FP8}},
howpublished={\url{https://huggingface.co/Qwen/Qwen3.5-397B-A17B-FP8}},
note={Published: February 2026. Accessed: September 2026},
year={2026},
month=feb,
}

@misc{qwen3527b_hf,
author={{Qwen Team}},
title={{Qwen3.5-27B-FP8}},
howpublished={\url{https://huggingface.co/Qwen/Qwen3.5-27B-FP8}},
note={Published: February 2026. Accessed: September 2026},
year={2026},
month=feb,
}

@misc{gemini15pro_api,
author={{Google}},
title={{Gemini 1.5 Pro}},
howpublished={\url{https://ai.google.dev/gemini-api/docs/models/gemini-1.5-pro}},
year={2024},
month=feb,
note={Published: February 2024. Accessed: September 2026},
}

@misc{gemini15flash_api,
author={{Google}},
title={{Gemini 1.5 Flash}},
howpublished={\url{https://ai.google.dev/gemini-api/docs/models/gemini-1.5-flash}},
year={2024},
month=may,
note={Published: May 2024. Accessed: September 2026},
}

@misc{gemini25pro_api,
author={{Google}},
title={{Gemini 2.5 Pro}},
howpublished={\url{https://blog.google/products/gemini/gemini-2-5-pro-updates/}},
year ={2025},
month=mar,
note ={Published: March 25, 2025. Accessed: September 2026},
}

@misc{gemini25flash_api,
author={{Google}},
title={{Gemini 2.5 Flash}},
howpublished={\url{https://developers.googleblog.com/en/start-building-with-gemini-25-flash/}},
year ={2025},
month=apr,
note ={Published: April 17, 2025. Accessed: September 2026},
}

@misc{gemini31pro_api,
author={{Google}},
title={{Gemini 3.1 Pro}},
howpublished={\url{https://ai.google.dev/gemini-api/docs/models/gemini-3.1-pro}},
year={2026},
month=feb,
note={Published: February 2026. Accessed: September 2026},
}

@misc{gemini37flash_api,
author={{Google}},
title={{Gemini 3.7 Flash}},
howpublished={\url{https://ai.google.dev/gemini-api/docs/models/gemini-3.7-flash}},
year={2026},
month=aug,
note={Published: August 2026. Accessed: September 2026},
}

@misc{openai2025codex,
author = {{OpenAI}},
title  = {Codex},
year   = {2025},
url    = {https://openai.com/codex/},
note   = {Accessed: September 2026}
}

@misc{google2025geminicli,
author = {{Google}},
title  = {Gemini {CLI}},
year   = {2025},
url    = {https://github.com/google-gemini/gemini-cli},
note   = {Accessed: September 2026}
}

@inproceedings{peng2025cweval,
  author       = {Jinjun Peng and
                  Leyi Cui and
                  Kele Huang and
                  Junfeng Yang and
                  Baishakhi Ray},
  title        = {CWEval: Outcome-driven Evaluation on Functionality and Security of
                  {LLM} Code Generation},
  booktitle    = {{IEEE/ACM} International Workshop on Large Language Models for Code,
                  LLM4Code@ICSE 2025, Ottawa, ON, Canada, May 3, 2025},
  pages        = {33--40},
  publisher    = {{IEEE}},
  year         = {2025},
  url          = {https://doi.org/10.1109/LLM4Code66737.2025.00009},
  doi          = {10.1109/LLM4CODE66737.2025.00009},
  bibsource    = {dblp computer science bibliography, https://dblp.org}
}

@inproceedings{cannavale2026smarter,
  title={Smarter, but Not Safer: An Empirical Analysis of the Functional-Security Gap in Evolving LLMs},
  author={Cannavale, Alfonso and Recupito, Gilberto and De Roover, Coen and Palomba, Fabio and De Lucia, Andrea},
  booktitle={International Conference on Software Maintenance and Evolution},
  pages={1--13},
  year={2026}
}

@article{wang2026solved,
  author       = {You Wang and
                  Michael Pradel and
                  Zhongxin Liu},
  title        = {Are "Solved Issues" in SWE-bench Really Solved Correctly?
                  An Empirical Study},
  journal      = {CoRR},
  volume       = {abs/2503.15223},
  year         = {2025},
  url          = {https://doi.org/10.48550/arXiv.2503.15223},
  doi          = {10.48550/arxiv.2503.15223},
  eprinttype   = {arXiv},
  eprint       = {2503.15223},
  bibsource    = {dblp computer science bibliography, https://dblp.org}
}

@article{chen2021evaluating,
  author={Chen, Mark and Tworek, Jerry and Jun, Heewoo and Yuan, Qiming and Pinto, Henrique Ponde De Oliveira and Kaplan, Jared and Edwards, Harri and Burda, Yuri and Joseph, Nicholas and Brockman, Greg and others},
  title        = {Evaluating Large Language Models Trained on Code},
  journal      = {CoRR},
  volume       = {abs/2107.03374},
  year         = {2021},
  url          = {https://doi.org/10.48550/arXiv.2107.03374},
  doi          = {10.48550/arxiv.2107.03374},
  eprinttype   = {arXiv},
  eprint       = {2107.03374},
  bibsource    = {dblp computer science bibliography, https://dblp.org}
}

@article{tihanyi2025secure,
  author       = {Norbert Tihanyi and
                  Tam{\'{a}}s Bisztray and
                  Mohamed Amine Ferrag and
                  Ridhi Jain and
                  Lucas C. Cordeiro},
  title        = {How secure is AI-generated code: a large-scale comparison of large
                  language models},
  journal      = {Empir. Softw. Eng.},
  volume       = {30},
  number       = {2},
  pages        = {47},
  year         = {2025},
  url          = {https://doi.org/10.1007/s10664-024-10590-1},
  doi          = {10.1007/S10664-024-10590-1},
  bibsource    = {dblp computer science bibliography, https://dblp.org}
}

@article{kharma2026security,
  author       = {Mohammed Kharma and
                  Soohyeon Choi and
                  Mohammed Alkhanafseh and
                  David Mohaisen},
  title        = {Security and Quality in LLM-Generated Code: {A} Multi-Language, Multi-Model
                  Analysis},
  journal      = {{IEEE} Trans. Dependable Secur. Comput.},
  volume       = {23},
  number       = {3},
  pages        = {7300--7314},
  year         = {2026},
  url          = {https://doi.org/10.1109/TDSC.2026.3672745},
  doi          = {10.1109/TDSC.2026.3672745},
  bibsource    = {dblp computer science bibliography, https://dblp.org}
}

@article{elgedawy2025ocassionally,
  author       = {Ran Elgedawy and
                  John Sadik and
                  Senjuti Dutta and
                  Anuj Gautam and
                  Konstantinos Georgiou and
                  Farzin Gholamrezae and
                  Fujiao Ji and
                  Kyungchan Lim and
                  Qian Liu and
                  Scott Ruoti},
  title        = {Ocassionally Secure: {A} Comparative Analysis of Code Generation Assistants},
  journal      = {CoRR},
  volume       = {abs/2402.00689},
  year         = {2024},
  url          = {https://doi.org/10.48550/arXiv.2402.00689},
  doi          = {10.48550/arxiv.2402.00689},
  eprinttype   = {arXiv},
  eprint       = {2402.00689},
  bibsource    = {dblp computer science bibliography, https://dblp.org}
}

@inproceedings{toth2024llms,
  author       = {Rebeka T{\'{o}}th and
                  Tam{\'{a}}s Bisztray and
                  L{\'{a}}szl{\'{o}} Erdodi},
  editor       = {Andrea Ceccarelli and
                  Mario Trapp and
                  Andrea Bondavalli and
                  Erwin Schoitsch and
                  Barbara Gallina and
                  Friedemann Bitsch},
  title        = {LLMs in Web Development: Evaluating LLM-Generated {PHP} Code Unveiling
                  Vulnerabilities and Limitations},
  booktitle    = {Computer Safety, Reliability, and Security. {SAFECOMP} 2024 Workshops
                  - DECSoS, SASSUR, TOASTS, and WAISE, Florence, Italy, September 17,
                  2024, Proceedings},
  series       = {Lecture Notes in Computer Science},
  volume       = {14989},
  pages        = {425--437},
  publisher    = {Springer},
  year         = {2024},
  url          = {https://doi.org/10.1007/978-3-031-68738-9\_34},
  doi          = {10.1007/978-3-031-68738-9\_34},
  bibsource    = {dblp computer science bibliography, https://dblp.org}
}

@misc{cwe,
author={{MITRE}},
title={{Common Weakness Enumeration (CWE)}},
howpublished={\url{https://cwe.mitre.org}},
note={Accessed: September 2026},
year={2026},
}

@misc{cwe_top25,
author={{MITRE}},
title={{CWE Top 25 Most Dangerous Software Weaknesses}},
howpublished={\url{https://cwe.mitre.org/top25/}},
note={Accessed: September 2026},
year={2026},
}

@inproceedings{tony2023llmseceval,
  author       = {Catherine Tony and
                  Markus Mutas and
                  Nicol{\'{a}}s E. D{\'{\i}}az Ferreyra and
                  Riccardo Scandariato},
  title        = {LLMSecEval: {A} Dataset of Natural Language Prompts for Security Evaluations},
  booktitle    = {20th {IEEE/ACM} International Conference on Mining Software Repositories,
                  {MSR} 2023, Melbourne, Australia, May 15-16, 2023},
  pages        = {588--592},
  publisher    = {{IEEE}},
  year         = {2023},
  url          = {https://doi.org/10.1109/MSR59073.2023.00084},
  doi          = {10.1109/MSR59073.2023.00084},
  bibsource    = {dblp computer science bibliography, https://dblp.org}
}

@inproceedings{siddiq2022securityeval,
author = {Siddiq, Mohammed Latif and Santos, Joanna C. S.},
title = {SecurityEval dataset: mining vulnerability examples to evaluate machine learning-based code generation techniques},
year = {2022},
isbn = {9781450394574},
publisher = {Association for Computing Machinery},
address = {New York, NY, USA},
url = {https://doi.org/10.1145/3549035.3561184},
doi = {10.1145/3549035.3561184},
booktitle = {Proceedings of the 1st International Workshop on Mining Software Repositories Applications for Privacy and Security},
pages = {29–33},
numpages = {5},
location = {Singapore, Singapore},
series = {MSR4P\&S 2022}
}

@article{bhatt2023purple,
  author       = {Manish Bhatt and
                  Sahana Chennabasappa and
                  Cyrus Nikolaidis and
                  Shengye Wan and
                  Ivan Evtimov and
                  Dominik Gabi and
                  Daniel Song and
                  Faizan Ahmad and
                  Cornelius Aschermann and
                  Lorenzo Fontana and
                  Sasha Frolov and
                  Ravi Prakash Giri and
                  Dhaval Kapil and
                  Yiannis Kozyrakis and
                  David LeBlanc and
                  James Milazzo and
                  Aleksandar Straumann and
                  Gabriel Synnaeve and
                  Varun Vontimitta and
                  Spencer Whitman and
                  Joshua Saxe},
  title        = {Purple Llama CyberSecEval: {A} Secure Coding Benchmark for Language
                  Models},
  journal      = {CoRR},
  volume       = {abs/2312.04724},
  year         = {2023},
  url          = {https://doi.org/10.48550/arXiv.2312.04724},
  doi          = {10.48550/ARXIV.2312.04724},
  eprinttype   = {arXiv},
  eprint       = {2312.04724},
  bibsource    = {dblp computer science bibliography, https://dblp.org}
}

@inproceedings{hajipour2024codelmsec,
  author       = {Hossein Hajipour and
                  Keno Hassler and
                  Thorsten Holz and
                  Lea Sch{\"{o}}nherr and
                  Mario Fritz},
  title        = {CodeLMSec Benchmark: Systematically Evaluating and Finding Security
                  Vulnerabilities in Black-Box Code Language Models},
  booktitle    = {{IEEE} Conference on Secure and Trustworthy Machine Learning, SaTML
                  2024, Toronto, ON, Canada, April 9-11, 2024},
  pages        = {684--709},
  publisher    = {{IEEE}},
  year         = {2024},
  url          = {https://doi.org/10.1109/SaTML59370.2024.00040},
  doi          = {10.1109/SATML59370.2024.00040},
  bibsource    = {dblp computer science bibliography, https://dblp.org}
}

@inproceedings{dai2025rethinking,
author = {Dai, Shih-Chieh and Xu, Jun and Tao, Guanhong},
title = {Rethinking the Evaluation of Secure Code Generation},
year = {2026},
isbn = {9798400720253},
publisher = {Association for Computing Machinery},
address = {New York, NY, USA},
url = {https://doi.org/10.1145/3744916.3773217},
doi = {10.1145/3744916.3773217},
booktitle = {Proceedings of the 2026 IEEE/ACM 48th International Conference on Software Engineering},
pages = {336–348},
numpages = {13},
location = {
},
series = {ICSE '26}
}

@inproceedings{vero2025baxbench,
  author       = {Mark Vero and
                  Niels M{\"{u}}ndler and
                  Victor Chibotaru and
                  Veselin Raychev and
                  Maximilian Baader and
                  Nikola Jovanovic and
                  Jingxuan He and
                  Martin T. Vechev},
  editor       = {Aarti Singh and
                  Maryam Fazel and
                  Daniel Hsu and
                  Simon Lacoste{-}Julien and
                  Felix Berkenkamp and
                  Tegan Maharaj and
                  Kiri Wagstaff and
                  Jerry Zhu},
  title        = {BaxBench: Can LLMs Generate Correct and Secure Backends?},
  booktitle    = {Forty-second International Conference on Machine Learning, {ICML}
                  2025, Vancouver, BC, Canada, July 13-19, 2025},
  series       = {Proceedings of Machine Learning Research},
  volume       = {267},
  publisher    = {{PMLR} / OpenReview.net},
  year         = {2025},
  url          = {https://proceedings.mlr.press/v267/vero25a.html},
  bibsource    = {dblp computer science bibliography, https://dblp.org}
}

@inproceedings{nie2026secodeplt,
  author       = {Yuzhou Nie and
                  Zhun Wang and
                  Yu Yang and
                  Ruizhe Jiang and
                  Yuheng Tang and
                  Xander Davies and
                  Yarin Gal and
                  Bo Li and
                  Wenbo Guo and
                  Dawn Song},
  editor       = {Danielle Belgrave and
                  Cheng Zhang and
                  Laura N. Montoya and
                  Hsuan{-}Tien Lin and
                  Razvan Pascanu and
                  Piotr Koniusz and
                  Marzyeh Ghassemi and
                  Nancy Chen and
                  Iv{\'{a}}n Vladimir Meza Ru{\'{\i}}z and
                  Arturo Loaiza{-}Bonilla},
  title        = {{SECODEPLT:} {A} Unified Benchmark for Evaluating the Security Risks
                  and Capabilities of Code GenAI},
  booktitle    = {Advances in Neural Information Processing Systems 38: Annual Conference
                  on Neural Information Processing Systems 2025, NeurIPS 2025, San Diego,
                  CA, USA, December 2-7, 2025 / Mexico City, Mexico, November 30 - December
                  5, 2025},
  year         = {2025},
  bibsource    = {dblp computer science bibliography, https://dblp.org},
  url={https://openreview.net/forum?id=vbr63iQsbv}
}

@inproceedings{lee2025secbench,
  author       = {Hwiwon Lee and
                  Ziqi Zhang and
                  Hanxiao Lu and
                  Lingming Zhang},
  editor       = {Danielle Belgrave and
                  Cheng Zhang and
                  Laura N. Montoya and
                  Hsuan{-}Tien Lin and
                  Razvan Pascanu and
                  Piotr Koniusz and
                  Marzyeh Ghassemi and
                  Nancy Chen and
                  Iv{\'{a}}n Vladimir Meza Ru{\'{\i}}z and
                  Arturo Loaiza{-}Bonilla},
  title        = {SEC-bench: Automated Benchmarking of {LLM} Agents on Real-World Software
                  Security Tasks},
  booktitle    = {Advances in Neural Information Processing Systems 38: Annual Conference
                  on Neural Information Processing Systems 2025, NeurIPS 2025, San Diego,
                  CA, USA, December 2-7, 2025 / Mexico City, Mexico, November 30 - December
                  5, 2025},
  year         = {2025},
  url          = {http://papers.nips.cc/paper\_files/paper/2025/hash/a9168f1c54e5147027f1e8cf83e1a775-Abstract-Conference.html},
  bibsource    = {dblp computer science bibliography, https://dblp.org}
}

@article{fu2025security,
  author       = {Yujia Fu and
                  Peng Liang and
                  Amjed Tahir and
                  Zengyang Li and
                  Mojtaba Shahin and
                  Jiaxin Yu and
                  Jinfu Chen},
  title        = {Security Weaknesses of Copilot-Generated Code in GitHub Projects:
                  An Empirical Study},
  journal      = {{ACM} Trans. Softw. Eng. Methodol.},
  volume       = {34},
  number       = {8},
  pages        = {218:1--218:34},
  year         = {2025},
  url          = {https://doi.org/10.1145/3716848},
  doi          = {10.1145/3716848},
  bibsource    = {dblp computer science bibliography, https://dblp.org}
}

@inproceedings{perry2023users,
  author       = {Neil Perry and
                  Megha Srivastava and
                  Deepak Kumar and
                  Dan Boneh},
  editor       = {Weizhi Meng and
                  Christian Damsgaard Jensen and
                  Cas Cremers and
                  Engin Kirda},
  title        = {Do Users Write More Insecure Code with {AI} Assistants?},
  booktitle    = {Proceedings of the 2023 {ACM} {SIGSAC} Conference on Computer and
                  Communications Security, {CCS} 2023, Copenhagen, Denmark, November
                  26-30, 2023},
  pages        = {2785--2799},
  publisher    = {{ACM}},
  year         = {2023},
  url          = {https://doi.org/10.1145/3576915.3623157},
  doi          = {10.1145/3576915.3623157},
  bibsource    = {dblp computer science bibliography, https://dblp.org}
}

@inproceedings{jimenez2024swebench,
  author       = {Carlos E. Jimenez and
                  John Yang and
                  Alexander Wettig and
                  Shunyu Yao and
                  Kexin Pei and
                  Ofir Press and
                  Karthik R. Narasimhan},
  title        = {SWE-bench: Can Language Models Resolve Real-world Github Issues?},
  booktitle    = {The Twelfth International Conference on Learning Representations,
                  {ICLR} 2024, Vienna, Austria, May 7-11, 2024},
  publisher    = {OpenReview.net},
  year         = {2024},
  url          = {https://openreview.net/forum?id=VTF8yNQM66},
  bibsource    = {dblp computer science bibliography, https://dblp.org}
}

@inproceedings{jain2024livecodebench,
  author       = {Naman Jain and
                  King Han and
                  Alex Gu and
                  Wen{-}Ding Li and
                  Fanjia Yan and
                  Tianjun Zhang and
                  Sida Wang and
                  Armando Solar{-}Lezama and
                  Koushik Sen and
                  Ion Stoica},
  title        = {LiveCodeBench: Holistic and Contamination Free Evaluation of Large
                  Language Models for Code},
  booktitle    = {The Thirteenth International Conference on Learning Representations,
                  {ICLR} 2025, Singapore, April 24-28, 2025},
  publisher    = {OpenReview.net},
  year         = {2025},
  url          = {https://openreview.net/forum?id=chfJJYC3iL},
  bibsource    = {dblp computer science bibliography, https://dblp.org}
}

@article{chen2023chatgpt,
  author       = {Lingjiao Chen and
                  Matei Zaharia and
                  James Zou},
  title        = {How is ChatGPT's behavior changing over time?},
  journal      = {CoRR},
  volume       = {abs/2307.09009},
  year         = {2023},
  url          = {https://doi.org/10.48550/arXiv.2307.09009},
  doi          = {10.48550/ARXIV.2307.09009},
  eprinttype   = {arXiv},
  eprint       = {2307.09009},
  bibsource    = {dblp computer science bibliography, https://dblp.org}
}

@article{liu2026robustness,
  author       = {Yugeng Liu and
                  Tianshuo Cong and
                  Zhengyu Zhao and
                  Michael Backes and
                  Yun Shen and
                  Yang Zhang},
  title        = {Robustness Over Time: Understanding Adversarial Examples' Effectiveness
                  on Longitudinal Versions of Large Language Models},
  journal      = {{IEEE} Trans. Inf. Forensics Secur.},
  volume       = {21},
  pages        = {3436--3449},
  year         = {2026},
  url          = {https://doi.org/10.1109/TIFS.2026.3672386},
  doi          = {10.1109/TIFS.2026.3672386},
  bibsource    = {dblp computer science bibliography, https://dblp.org}
}

@article{liang1986longitudinal,
 ISSN = {00063444},
 URL = {http://www.jstor.org/stable/2336267},
 author = {Kung-Yee Liang and Scott L. Zeger},
 journal = {Biometrika},
 number = {1},
 pages = {13--22},
 publisher = {[Oxford University Press, Biometrika Trust]},
 title = {Longitudinal Data Analysis Using Generalized Linear Models},
 urldate = {2026-10-02},
 volume = {73},
 year = {1986}
}

@inproceedings{he2024instructiontuningsecurecode,
  author       = {Jingxuan He and
                  Mark Vero and
                  Gabriela Krasnopolska and
                  Martin T. Vechev},
  editor       = {Ruslan Salakhutdinov and
                  Zico Kolter and
                  Katherine A. Heller and
                  Adrian Weller and
                  Nuria Oliver and
                  Jonathan Scarlett and
                  Felix Berkenkamp},
  title        = {Instruction Tuning for Secure Code Generation},
  booktitle    = {Forty-first International Conference on Machine Learning, {ICML} 2024,
                  Vienna, Austria, July 21-27, 2024},
  series       = {Proceedings of Machine Learning Research},
  volume       = {235},
  pages        = {18043--18062},
  publisher    = {{PMLR} / OpenReview.net},
  year         = {2024},
  url          = {https://proceedings.mlr.press/v235/he24k.html},
  bibsource    = {dblp computer science bibliography, https://dblp.org}
}

@inproceedings{qi2024fine,
  author       = {Xiangyu Qi and
                  Yi Zeng and
                  Tinghao Xie and
                  Pin{-}Yu Chen and
                  Ruoxi Jia and
                  Prateek Mittal and
                  Peter Henderson},
  title        = {Fine-tuning Aligned Language Models Compromises Safety, Even When
                  Users Do Not Intend To!},
  booktitle    = {The Twelfth International Conference on Learning Representations,
                  {ICLR} 2024, Vienna, Austria, May 7-11, 2024},
  publisher    = {OpenReview.net},
  year         = {2024},
  url          = {https://openreview.net/forum?id=hTEGyKf0dZ},
  bibsource    = {dblp computer science bibliography, https://dblp.org}
}

@article{recupito2024technical,
  author       = {Gilberto Recupito and
                  Fabiano Pecorelli and
                  Gemma Catolino and
                  Valentina Lenarduzzi and
                  Davide Taibi and
                  Dario Di Nucci and
                  Fabio Palomba},
  title        = {Technical debt in AI-enabled systems: On the prevalence, severity,
                  impact, and management strategies for code and architecture},
  journal      = {J. Syst. Softw.},
  volume       = {216},
  pages        = {112151},
  year         = {2024},
  url          = {https://doi.org/10.1016/j.jss.2024.112151},
  doi          = {10.1016/J.JSS.2024.112151},
  bibsource    = {dblp computer science bibliography, https://dblp.org}
}

@inproceedings{elahi2011security,
  author       = {Golnaz Elahi and
                  Eric S. K. Yu and
                  Tong Li and
                  Lin Liu},
  title        = {Security Requirements Engineering in the Wild: {A} Survey of Common
                  Practices},
  booktitle    = {Proceedings of the 35th Annual {IEEE} International Computer Software
                  and Applications Conference, {COMPSAC} 2011, Munich, Germany, 18-22
                  July 2011},
  pages        = {314--319},
  publisher    = {{IEEE} Computer Society},
  year         = {2011},
  url          = {https://doi.org/10.1109/COMPSAC.2011.48},
  doi          = {10.1109/COMPSAC.2011.48},
  bibsource    = {dblp computer science bibliography, https://dblp.org}
}

@inproceedings{kwon2023vllm,
  author       = {Woosuk Kwon and
                  Zhuohan Li and
                  Siyuan Zhuang and
                  Ying Sheng and
                  Lianmin Zheng and
                  Cody Hao Yu and
                  Joseph Gonzalez and
                  Hao Zhang and
                  Ion Stoica},
  editor       = {Jason Flinn and
                  Margo I. Seltzer and
                  Peter Druschel and
                  Antoine Kaufmann and
                  Jonathan Mace},
  title        = {Efficient Memory Management for Large Language Model Serving with
                  PagedAttention},
  booktitle    = {Proceedings of the 29th Symposium on Operating Systems Principles,
                  {SOSP} 2023, Koblenz, Germany, October 23-26, 2023},
  pages        = {611--626},
  publisher    = {{ACM}},
  year         = {2023},
  url          = {https://doi.org/10.1145/3600006.3613165},
  doi          = {10.1145/3600006.3613165},
  bibsource    = {dblp computer science bibliography, https://dblp.org}
}

@article{hou2024large,
  author       = {Xinyi Hou and
                  Yanjie Zhao and
                  Yue Liu and
                  Zhou Yang and
                  Kailong Wang and
                  Li Li and
                  Xiapu Luo and
                  David Lo and
                  John Grundy and
                  Haoyu Wang},
  title        = {Large Language Models for Software Engineering: {A} Systematic Literature
                  Review},
  journal      = {{ACM} Trans. Softw. Eng. Methodol.},
  volume       = {33},
  number       = {8},
  pages        = {220:1--220:79},
  year         = {2024},
  url          = {https://doi.org/10.1145/3695988},
  doi          = {10.1145/3695988},
  bibsource    = {dblp computer science bibliography, https://dblp.org}
}

@article{xia2024evoeval,
  author       = {Chunqiu Steven Xia and
                  Yinlin Deng and
                  Lingming Zhang},
  title        = {Top Leaderboard Ranking = Top Coding Proficiency, Always? EvoEval:
                  Evolving Coding Benchmarks via {LLM}},
  journal      = {CoRR},
  volume       = {abs/2403.19114},
  year         = {2024},
  url          = {https://doi.org/10.48550/arXiv.2403.19114},
  doi          = {10.48550/arXiv.2403.19114},
  eprinttype   = {arXiv},
  eprint       = {2403.19114},
  bibsource    = {dblp computer science bibliography, https://dblp.org}
}

@article{pearce2025asleep,
  author       = {Hammond Pearce and
                  Baleegh Ahmad and
                  Benjamin Tan and
                  Brendan Dolan{-}Gavitt and
                  Ramesh Karri},
  title        = {Asleep at the Keyboard? Assessing the Security of GitHub Copilot's
                  Code Contributions},
  journal      = {Commun. {ACM}},
  volume       = {68},
  number       = {2},
  pages        = {96--105},
  year         = {2025},
  url          = {https://doi.org/10.1145/3610721},
  doi          = {10.1145/3610721},
  bibsource    = {dblp computer science bibliography, https://dblp.org}
}

@inproceedings{khoury2023secure,
  author       = {Rapha{\"{e}}l Khoury and
                  Anderson R. Avila and
                  Jacob Brunelle and
                  Baba Mamadou Camara},
  title        = {How Secure is Code Generated by ChatGPT?},
  booktitle    = {{IEEE} International Conference on Systems, Man, and Cybernetics,
                  {SMC} 2023, Honolulu, Oahu, HI, USA, October 1-4, 2023},
  pages        = {2445--2451},
  publisher    = {{IEEE}},
  year         = {2023},
  url          = {https://doi.org/10.1109/SMC53992.2023.10394237},
  doi          = {10.1109/SMC53992.2023.10394237},
  bibsource    = {dblp computer science bibliography, https://dblp.org}
}

@inproceedings{siddiq2024quality,
  author       = {Mohammed Latif Siddiq and
                  Lindsay Roney and
                  Jiahao Zhang and
                  Joanna C. S. Santos},
  editor       = {Diomidis Spinellis and
                  Alberto Bacchelli and
                  Eleni Constantinou},
  title        = {Quality Assessment of ChatGPT Generated Code and their Use by Developers},
  booktitle    = {21st {IEEE/ACM} International Conference on Mining Software Repositories,
                  {MSR} 2024, Lisbon, Portugal, April 15-16, 2024},
  pages        = {152--156},
  publisher    = {{ACM}},
  year         = {2024},
  url          = {https://doi.org/10.1145/3643991.3645071},
  doi          = {10.1145/3643991.3645071},
  bibsource    = {dblp computer science bibliography, https://dblp.org}
}

@article{sergeyuk2026human,
  author       = {Agnia Sergeyuk and
                  Ilya Zakharov and
                  Ekaterina Koshchenko and
                  Maliheh Izadi},
  title        = {Human-AI experience in integrated development environments: a systematic
                  literature review},
  journal      = {Empir. Softw. Eng.},
  volume       = {31},
  number       = {3},
  pages        = {55},
  year         = {2026},
  url          = {https://doi.org/10.1007/s10664-025-10793-0},
  doi          = {10.1007/S10664-025-10793-0},
  bibsource    = {dblp computer science bibliography, https://dblp.org}
}

@article{kemerer1999empirical,
  author       = {Chris F. Kemerer and
                  Sandra Slaughter},
  title        = {An Empirical Approach to Studying Software Evolution},
  journal      = {{IEEE} Trans. Software Eng.},
  volume       = {25},
  number       = {4},
  pages        = {493--509},
  year         = {1999},
  url          = {https://doi.org/10.1109/32.799945},
  doi          = {10.1109/32.799945},
  bibsource    = {dblp computer science bibliography, https://dblp.org}
}

@inproceedings{afrin2025quantization,
  author       = {Saima Afrin and
                  Bowen Xu and
                  Antonio Mastropaolo},
  title        = {Is Quantization a Deal-Breaker? Empirical Insights From Large Code
                  Models},
  booktitle    = {{IEEE} International Conference on Software Maintenance and Evolution,
                  {ICSME} 2025, Auckland, New Zealand, September 7-12, 2025},
  pages        = {1--13},
  publisher    = {{IEEE}},
  year         = {2025},
  url          = {https://doi.org/10.1109/ICSME64153.2025.00049},
  doi          = {10.1109/ICSME64153.2025.00049},
  bibsource    = {dblp computer science bibliography, https://dblp.org}
}

@article{vallecillos2026smaller,
  author       = {Fernando Vallecillos Ruiz and
                  Giordano d'Aloisio and
                  Max Hort and
                  Luca Traini and
                  Antinisca Di Marco and
                  Leon Moonen},
  title        = {Smaller Models, Unexpected Costs: Trade-offs in {LLM} Quantization
                  for Automated Program Repair},
  journal      = {CoRR},
  volume       = {abs/2606.27205},
  year         = {2026},
  url          = {https://doi.org/10.48550/arXiv.2606.27205},
  doi          = {10.48550/ARXIV.2606.27205},
  eprinttype   = {arXiv},
  eprint       = {2606.27205},
  bibsource    = {dblp computer science bibliography, https://dblp.org}
}

\end{document}